\documentclass[universe,review,accept,moreauthors,pdftex]{Definitions/mdpi} 

\firstpage{1} 
\pubvolume{1}
\issuenum{1}
\articlenumber{0}
\pubyear{2021}
\copyrightyear{2021}
\externaleditor{Academic Editor: Norma G. Sanchez}
\datereceived{19 May 2021} 
\dateaccepted{25 August 2021} 
\datepublished{9 September 2021} 
\hreflink{https://doi.org/} 

\usepackage{xcolor}
\usepackage{enotez}

\Title{Fundamental Properties of the Dark and the Luminous Matter from the Low Surface Brightness Discs}

\TitleCitation{Fundamental Properties of the Dark and the Luminous Matter from the Low Surface Brightness Discs}

\Author{{Paolo Salucci} *$^{,\dagger,\ddagger}$\orcidA{} and Chiara di Paolo $^{\ddagger}$ }

\AuthorNames{Paolo Salucci and Chiara di Paolo}

\AuthorCitation{Salucci, P.; di Paolo, C.}

\address[1]{INFN Sezione di Trieste, Via A. Valerio 2, 34127 Trieste, Italy; cdipaolo@sissa.it}

\corres{\hangafter=1 \hangindent=1em \hspace{-1em} Correspondence: salucci@sissa.it}

\firstnote{\hangafter=1 \hangindent=1em \hspace{-1em} Current address: SISSA, International School for Advanced Studies, Via Bonomea 265, 34136 Trieste, Italy.} 
\secondnote{\hangafter=1 \hangindent=1em \hspace{-1em} These authors contributed equally to this work.}

\abstract{Dark matter (DM) is one of the biggest mystery in the Universe. In this review, we start reporting the evidences for this elusive component and discussing about the proposed particle candidates and scenarios for such phenomenon. Then, we focus on recent results obtained for rotating disc galaxies, in particular for low surface brightness (LSB) galaxies. The main observational properties related to the baryonic matter in LSBs, investigated over the last decades, are briefly recalled. Next, these galaxies are analyzed by means of the mass modelling of their rotation curves both individual and stacked. The latter analysis, via the universal rotation curve (URC) method, results really powerful in giving a global or universal description of the properties of these objects. We report the presence in LSBs of scaling relations among their structural properties that result comparable with those found in galaxies of different morphologies. All this confirms, in disc systems, the existence of a strong entanglement between the luminous matter (LM) and the dark matter (DM).
Moreover, we report how in LSBs the tight relationship between their radial gravitational accelerations $g$ and their baryonic components $g_b$ results to depend also on the stellar disk length scale and the radius at which the two accelerations have been measured. 
LSB galaxies strongly challenge the $\Lambda $CDM scenario with the relative collisionless dark particle and, alongside with the non-detection of the latter, contribute to guide us towards a new scenario for the DM phenomenon.}


\begin{document}

\section{Introduction}
\label{introduction}
By means of (radio)-telescopes it is possible to observe the ``light'' emitted by stars, dust, and gas in galaxies  but this is only the tip of an iceberg of their total mass. 
More generally, according to the latest observational data, the mass energy of the Universe contains
only $\sim$5\% in baryonic ordinary matter, $\sim$27\% in dark matter, and $\sim$68\% in dark energy (e.g., \citep{Ade_2014, Aghanim_2018}).

{\it {Dark matter}} (DM) is a type of matter put forward in order to account for effects on the luminous matter (LM) that appear to arise from an invisible massive component. 
In detail, the existence and the properties of the dark matter can be inferred from its gravitational effects on the luminous matter and radiation and from the properties of the large-scale structure of the Universe. 
Astrophysicists have hypothesised the existence of such ``dark matter'' as result of severe discrepancies between the distribution of the gravitating mass of large cosmological objects and that of the ``luminous matter'' that they contain (stars, gas, and dust). The presence of dark matter emerges in the rotational speeds of \mbox{galaxies~\citep{Faber_1979,Trimble_1987, Rubin_1980, Bosma_1, Bosma_1981},} in the gravitational lensing of background objects \citep{oka}, in the extraordinary properties of the bullet cluster \citep{Clowe_2004}, in the temperature distribution of hot gas around galaxies and clusters of galaxies \citep{Rees_1977, Cavaliere_1978} and in the pattern of anisotropies of the cosmic microwave ground (CMB) radiation \citep{ Ade_2016} that implies that about five-sixths of the total matter does not interact significantly with ordinary standard model particles. Furthermore, the theory of big bang nucleosynthesis (BBN), which accurately predicts the observed abundance of the light elements $^2$D and $^4$He, indicates that the majority of matter in the Universe cannot be made by BBN baryons \citep{Persic_1992, Copi_1995, Nicastro_2018}. In agreement with this, accurate gravitational microlensing measurements have shown that only a small fraction of the dark matter in the Milky Way could be hidden in (maybe primordial) dark compact objects composed of ordinary (baryonic) matter emitting little or no electromagnetic \mbox{radiation~\citep{Alcock_2000,Tisserand_2007, Wyrzykowski_2011}.} All this implies a non-baryonic and non-standard Model (SM) nature for the dark matter particle. The existence of such a particle, necessarily beyond the standard model (SM), is hoped or assumed to solve pressing problems inside the SM itself or to expand the knowledge of particle physics into new territories.
It is well-known that the DM phenomenon is framed in the currently most favoured $\Lambda$-cold dark matter ($\Lambda$CDM) scenario \citep{Kolb_Turner_1990, Mukhanov_2005, Ellis_2012} where the non-relativistic DM can be described as a collisionless fluid whose particles interact (almost) only gravitationally and very weakly with the SM particles (\citep{Jungman_1996, Bertone}).
However, despite the evidences on its existence, this mysterious component of the Universe is not yet characterised. In addition to several observational issues that complicate its identification, the search for such a particle, performed by a variety of methods, despite being in the past 20 years one of the major efforts in (astro)particle physics, has resulted unsuccessful \mbox{(e.g.,~\citep{Bertone_2005, Arcadi_2018}).} However, we maintain here the scenario of particle dark matter, in that, in addition to successfully accounting for the very existence of virialized objects as galaxies, is able to cope with their formation process and with the large scale properties of the entire Universe, all goals that seem unreachable for alternative scenarios (as MOND \citep{Milgrom_1983}, F(R)-gravity and scalar–tensor gravity \citep{Capozziello_2011}). Finally, we think that the dark particle scenario is not obliged to follow the paradigm according to which the particle must be the simplest, the most elegant, the most theoretically favoured and the most expected beyond SM (see~\citep{Salturini}). 

In the past 25 years, the dark matter properties at galactic scales have progressively increased their importance within the puzzle of the dark matter phenomenon. On the observational side, the special importance of certain families of galaxies is well known. In detail, dwarf spheroidals are a primary target to study the DM phenomenon (e.g., \citep{Lokas}). They have the advantage to be dark matter dominated at any radius and to lay in a dark halo mass range in which the discrepancy between the $\Lambda$CDM scenario predictions and the actual observations are expected to be very apparent. Finally, as regards the indirect detection of the DM particle, they are the nearest objects to search for. Noticeably,  the kinematics of DM-dominated dwarf disks also provide us with valuable information on the properties of the DM halos \citep{Karukes_2017}.

This review is focused on recent investigations on low surface brightness (LSB) galaxies and, namely, on their DM distribution, its relation with the luminous matter distribution and the implications on the DM mystery. In other words, it is centered on the structural properties of the DM and the LM in LSBs, galaxies that belong to the family of discs, i.e., rotating objects with a rather simple kinematics.

\begin{figure}[H]
\includegraphics[width=0.65\textwidth,angle=0,clip=true]{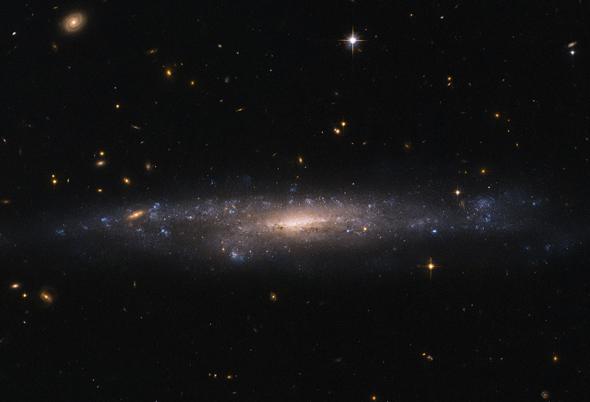}
\caption{A typical LSB galaxy (UGC 477). {Credits}: ESA/Hubble and NASA.}
\label{LSB_location}
\end{figure}

These systems emit an amount of light per area smaller than normal spirals (see \mbox{Figure~\ref{LSB_location}),} in fact, by definition, they have a face-on central surface {brightness}\linebreak $\gtrsim$23$\, \rm mag \,arcsec^{-2}$ in the B band \citep{Impey_1997}. They are more isolated than the high surface brightness (HSB) galaxies (e.g., \citep{Bothun_1993}) and are characterised by very low star formation rates (SFR) ($\lesssim$0.1 $\rm\ M_{\odot}\, yr^{-1}$) and SFR surface densities ( $\lesssim$0.001 $\rm \, M_{\odot} \, yr^{-1}kpc^{-2}$) (e.g., \citep{Das_2009}). They also have particular colours, metallicities and gas fractions (e.g., \citep{vanderHulst_1993}). 
Radio synthesis observations show that these objects have very massive and very extended gaseous discs  although with surface densities not higher than $\simeq$5 $\ M_{\odot}/pc^2$ and $M_{HI}/L_B$ ratios high 
up to $\sim$50 (e.g.,~\citep{vanderHulst_1993}), with $M_{HI}$ the mass of the HI disc. 
Furthermore, inside their optical radius $R_{opt}$,\endnote{That encloses 83\% of the total disk light} LSBs are largely dominated by DM, as shown by the analysis of their: (1)~Tully--Fisher relation (e.g., \citep{Zwaan_1995}), (2) individual (e.g., \citep{deBlok_2001, deBlok_2002}) and stacked \citep{DiPaolo_2019} rotation curves (RCs). 
Given their unique peculiarities that include: a very large extension, low surface densities in the stellar and gaseous disks so as in the 2-D projected DM surface density: $2 \,\int^\infty_0 \ \rho_{DM}(R,z) \, dz $, large amounts of DM and extremely low star formation rates, LSBs are very promising systems to help resolving the dark matter puzzle, being very different cosmic laboratories with respect to normal spirals, dwarf disks, and dwarf, normal, and giant ellipticals.
 
The topic of this review, focused on the {{properties}} of dark and luminous matter in LSB disc galaxies and their {\bf implications} for the DM mystery, is related to several other topics in astrophysics, cosmology, astroparticle physics, and physics of the elementary particles. In relation with the latter we suggest: 
``Galaxy Disks'' \citep{van_der_Kruit_2011}, 
``The Standard Cosmological Model: Achievements and Issues'' \citep{Ellis_2018},
``WIMP dark matter candidates and searches---current status and future prospects'' \citep{Roszkowski_2018}, 
``Status of dark matter in the universe''~\citep{Freese_2017}. A general review of the issue of the DM in galaxies can be found in \citep{Salucci_2019}.
Furthermore, in the next sections, we will indicate a number of works that extend the content of the present~review.

\section{DM Phenomenon in the Particles Framework}
\label{DM_candidates}
After accepting the existence of dark matter,  a spontaneous question arises: what is its nature? Several possibilities have been proposed. Even though this review is on the ``astrophysics'' side of the dark matter, it is still
necessary to consider the elementary particle (EP) side of the DM phenomenon. The involved elementary particle is likely extremely long-lived and stable, with a lifetime comparable to the age of the Universe, as suggested by the large cosmic abundance of DM that must have been generated very early in the history of the Universe (though there are exceptions (e.g., \citep{Sarkar_2015})) and survived mostly unchanged until today,
(e.g., \citep{Kolb_Turner_1990}) at least outside of the innermost regions of galaxies. In the latter, in fact,
very often the baryonic matter strongly dominates the gravitational potential, so that the fate, over the Gyrs, of the dark particles cannot be easily tracked~down. 

In the following, the most favoured dark particle candidates, whose actual presence in LSBs can be tested, are shortly introduced. For a complete discussion of the various DM particle models and existing constraints, see, e.g., \citep{Bergstrom_2000, Bertone_2005, Garrett_2011, Bauer_2017, Profumo_2017}.

\subsection{Weakly Interacting Massive Particles (WIMPs)}
\label{WIMP}
Weakly interacting massive particles (WIMPs) are particles that are thought to interact via gravity and via an interaction beyond the SM as weak as (or weaker than) the weak nuclear interaction (i.e., with a cross {section} $\sigma \lesssim 10^{-26}$ cm$^2$). These particles are {\bf collisionless} and, therefore, their dynamical evolution can be well investigated by N-body simulations. In more detail, WIMPs perfectly interpret the model of a relic particle with mass $m_{_{\rm WIMP}}$ coming from the early Universe, when all particles were in a state of thermal equilibrium. For temperatures $T \gg m_{_{\rm WIMP}}$, existing in the early Universe, the dark matter particle and its antiparticle are both forming from (and annihilating into) lighter particles of the standard model ($DM + DM \rightleftharpoons SM + SM $). As the Universe expands and cools ($ T \lesssim m_{_{\rm WIMP}} $), the average thermal energy of these lighter particles decreases and eventually becomes too small to form a dark matter particle--antiparticle pair. The annihilation of the dark matter particle--antiparticle pairs 
($DM + DM \Rightarrow SM + SM $) continues and the number density of dark matter particles begins to decrease exponentially ($\propto exp[-m_{_{\rm WIMP}}/T]$). Then, the number density becomes so low that the dark matter particle--antiparticle interaction stops, and the ratio between dark matter and photon densities ``freezes-out'', i.e., remains constant as the Universe continues to expand. The ‘freeze-out’ time occurs when the annihilation rate $\Gamma$ is on the order of the Hubble rate: $ \Gamma \sim n_{\rm DM}<\sigma v>\sim H^{-1}$, where $n_{DM}$ is the DM number density and $<\sigma v >$ is the velocity-averaged cross section.

A particle in the 10 GeV to 10 TeV mass range that interacts via the electroweak force with a typical self-annihilation cross section of $\langle \sigma v\rangle \simeq 3\times 10^{-26}$ {cm}$^{3}$ {s}$^{-1}$ implies a relic density similar to the cosmological matter density $\Omega_m \rho_c\sim 0.3 \times 10^{-29}$ g/cm$^3$ \citep{Steigman}. 

Noticeably, the resulting freeze out velocity is much smaller than $c$, so that the dark particles can be considered ``cold'' and initially at rest with respect to each other. Supersymmetric extensions of the SM of particle physics readily predict a particle with the properties described above and with the in-built ``WIMP miracle'' \citep{Steigman_1985, Kolb_Turner_1990, Jungman_1996, Munoz_2017}. However, it is worthwhile to anticipate that such extensions are almost ruled out by the fact that LHC has
not detected charginos or neutralinos from the decay of B-mesons \citep{Aaij_2012,Bechtle_2012}.

\begin{figure}[H]
\smallskip
\includegraphics[width=0.61\textwidth,angle=0,clip=true]{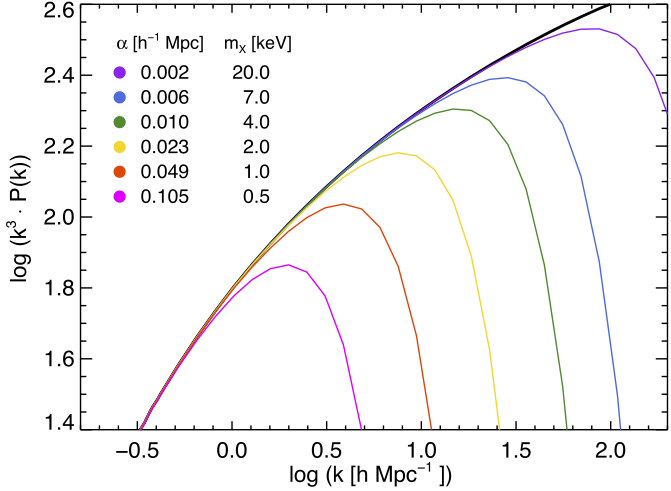}
\caption{Linear power spectra for $\Lambda$CDM (black) and $\Lambda$WDM (coloured) scenarios according to their thermal relic mass and damping scale $\alpha$. For a mass of 2 keV the power spectrum has a cut-off at galactic scales. 
Image reproduced {from} \citep{Kennedy_2014}.}
\label{PWS}
\end{figure}

Because of their large mass, WIMPs move relatively slow: they are cold dark matter (CDM) particles, characterised by non-relativistic velocities at the decoupling time.
Such low velocities cannot overcome those originating from the mutual gravitational attraction and, therefore, WIMPs clump together, from small structures to the largest ones (bottom-up scenario). They have a particular power spectrum of perturbations (see Figure~\ref{PWS}) with Gaussian initial conditions that are independent of the following evolution of the density~perturbations. 

WIMPs are described here in details in that they are currently considered as the reference DM particles. This is earned from the simplicity of the scenario, the above ``miracle'', the relation with SuSy and the agreement between the WIMP predictions and a number of cosmological large scale observations. It is well known that N-body simulations, performed in this particle scenario framed into the currently favoured cosmological concordance model (i.e., the Plank cosmology), show that the evolution of the inbuilt density perturbations give rise to virialized DM halos with a rather universal spherically averaged density profile. A successful and much preferred fit of the latter is the well known Navarro--Frenk--White (NFW) profile 
$\rho_{\rm NFW}(r)$ \citep{Navarro_1997}:
\begin{eqnarray}
\label{NFW_profile}
\rho_{\rm NFW}(r)= \frac{\rho_s}{(r/r_s)(1+r/r_s)^2} =\frac{M_{vir}}{4\pi R_{vir}} 
\frac{c^2 g(c)}{\tilde{x}(1+c\tilde{x})^2} \
\end{eqnarray} where the density $\rho_s$ and the scale radius $r_s$ are parameters which vary from halo to
halo in a strongly correlated way (e.g., \citep{Wechsler_2006}), $\tilde{x} =r/R_{vir}$ and the concentration parameter $c$ is defined as: $c \equiv r_s/R_{vir}$, with $R_{vir}$ the virial radius,\endnote{The virial radius $R_{vir}$ is defined as the radius at which the DM mass inside it is 200 times the critical density of the Universe times the volume inside this radius.} which encloses the whole mass of the dark halo and: $g(c)=[ln(1+c)-c/(1+c)]^{-1}$. At redshift zero this parameter results a weak function of the halo mass \citep{Klypin_2011}. One must highlight the inner {\it cuspy} profile ($\rho_{\rm NFW}\propto r^{-a}$, with $a= -1$) which is a decisive feature of the $\Lambda\rm  CDM$ dark halos.

Anticipating the content of next section we must point out that this scenario has serious issues: the WIMP particle has not been detected till now (see Section \ref{Experiments}) and it is strongly challenged by the DM astrophysical properties at small scales (see, e.g., \citep{Naab_2017, Bullock_2017, Salucci_2019} and Section \ref{Issues_with_the_main_DM_scenario}). It is, therefore, mandatory to consider other scenarios for the dark particle also supported by theoretical considerations.

\subsection{Scalar Fields and Fuzzy Dark Matter}
\label{Fuzzy}
Ultralight axion (ULA) with $m_{\psi} \sim 10^{-22} \,$eV is a scalar field particularly interesting in DM astrophysics \citep{Weinberg_1978, Hu_2000, Ringwald_2012, Hui_2017, Bernal_2017}. 
On theoretical grounds, it is worth recalling that the axion is introduced in order to solve the strong CP problem in particle physics (e.g., see \citep{Duffy_2009}). Furthermore, other scalar fields as axion-like particles were introduced, motivated by string theory \citep{Kane_2015}. These scalars are required to be non-relativistic and abundantly produced in very early Universe and to be (subsequently or always) decoupled from ordinary matter. 
These particles, at large scales, mimic the behaviour of the CDM particles, but, at small galactic scales, where the inter-particle distance is much smaller than their de Broglie wave length, move collectively as a wave and their equation of state can lead the DM density to cored configurations like those observed. We have, then, the {\bf fuzzy DM} scenario with the particles behaving as Bose--Einstein condensates (BEC).
As a reference starting point, the ULA-DM halo density profile assumes the following profile (\citep{Schive_2014} and Figure \ref{LSB_location} therein):
\begin{eqnarray}
\label{Fuzzy_eq}
\rho_{\rm ULA} (r)= \frac{1.9 \, a^{-1} (m_{\psi}/10^{-23} eV)^{-2} (r_c/ kpc)^{-4}}{[1+ 9.1 \times 10^{-2} (r/r_c)^2]^8} M_{\odot} pc^{-3},
\end{eqnarray}
where $a$ is the cosmic scale factor ($a(z=0)=1$), $m_{\psi}$ is particle mass and $r_c$ is the core radius defined as the radius at which the density drops to a value one-half of its peak value.

\subsection{Self-Interacting Dark Matter (SIDM)}
\label{SIDM}
\begin{figure}[H]
\includegraphics[width=0.65\textwidth,angle=0,clip=true]{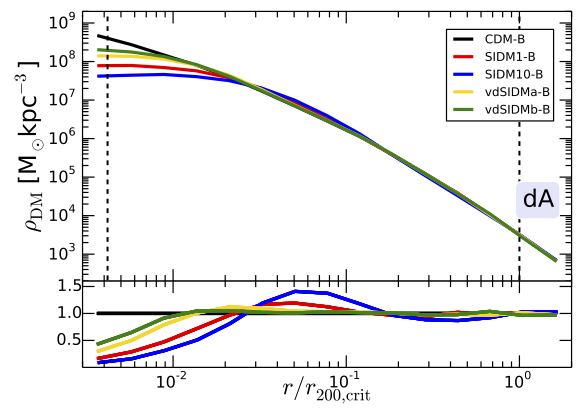}
\caption{DM halo profile for collisionless standard $\Lambda$CDM (black) and for the self-interacting DM scenario (SIDM) with different cross-section (coloured), see \citep{Vogelsberger_2014}. 
}
\label{SIDM_profile}
\end{figure}

One can assume that the dark matter particles are subject to self-interactions and this scenario could resolve a number of conflicts, at galactic scales, between observations and N-body simulations (of cold collisionless dark matter) \citep{Spergel_2000}. According to this scenario, the dark matter is self-interacting with a large scattering cross-section but negligible annihilation or dissipation. The large scattering cross-section may be due to strong, short-range interactions, similar to neutron--neutron scattering at low-energies, or to weak interactions mediated by the exchange of light particles with {masses} $\sim$0.5 $\, {\rm keV} (m_p/{\rm GeV})$, where $m_p$ is the mass of the particle \citep{Spergel_2000}.
In such a scenario, the DM particles scatter elastically with each other and, inside the dense inner region of the halo, get heated and expelled out it, a~process that reduce the density of the inner regions of the dark halos. In short, the original cusped profile is transformed in a cored one. Let us stress that the collision rate is negligible during the early Universe when the cosmological structures were formed. Therefore, this model retains the large-scale successes of the $\Lambda$CDM scenario, especially with a velocity dependent cross section that might help reconciling the properties of the predicted halos with those observed (see \citep{Zavala_2013, Tulin_2013, Bellazzini_2013, Boddy_2014, Vogelsberger_2014, Elbert_2015, Kaplinghat_2015}). 
For a dwarf galaxy, the resulting SIDM profiles, with a core radius that can reach $3 \times 10^{-2} R_{vir}$, are shown in Figure \ref{SIDM_profile} in relation with the value of the cross section. Remarkably, once we choose a value for this quantity, unless it is strongly velocity dependent, the dimensions of the resulting core radii are similar in halos of different masses.
%

\subsection{Sterile Neutrino: Warm Dark Matter Particle}
\label{WDM}
\begin{figure}[H]
\hspace{-10pt}\includegraphics[width=0.60\textwidth,height=0.44\textwidth]{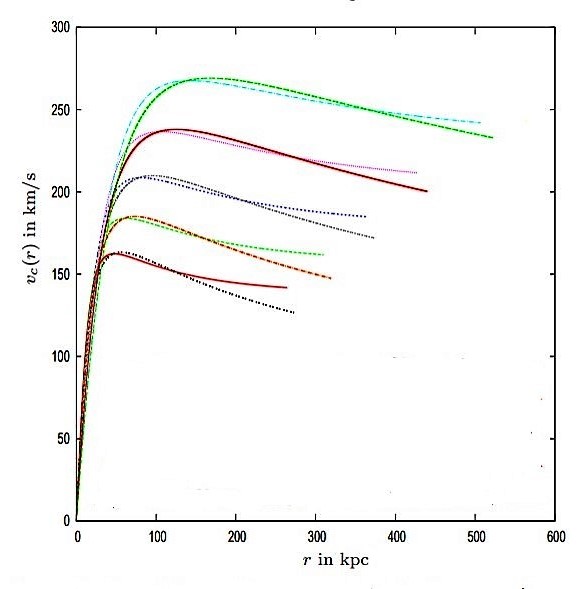}
\caption{{The} universal rotation curve of spirals compared with the predictions of the 2 keV WDM scenario (that flattens at outermost radii) \citep{de_Vega_2014}.}
\label{WDM_profile}
\end{figure}

The sterile neutrino is a lepton particle beyond the standard model of particle physics. It is thought to interact only via gravity and not via other fundamental interactions \mbox{(e.g., \citep{Drewes_2013,Adhi, Boyarsky_2019})}. The existence of this particle is motivated by arguments on the chirality of fermions and on the possibility to explain in a natural way the small active neutrino masses through the seesaw mechanism (e.g., \citep{Asaka_2005, Ma_2006}).
The sterile neutrino in the keV mass range (e.g., \citep{Drewes_2013, Naumov_2019}) is a DM particle candidate able to overcome the problems at small scales of the CDM scenario (\cite{dvs2012p}, for a review: \citep{Adhi}). It is classified as warm dark matter (WDM) particle and can be created in the early Universe (\citep{Dodelson_1994, Shi_1999, Kusenko_2009}); it decouples from the cosmological plasma when still mildly relativistic. WDM particles with masses of the keV have a power spectrum with a cut off at galactic scales that eliminates the overabundance of halos at such masses, plaguing the collisionless $\Lambda$CDM scenario (see Figure \ref{PWS}). Moreover, by taking into account the fermionic nature of this particle, one realises that it deals also with the cusp issue. In fact, for masses of {about} $\sim$keV, the particle de-Broglie scale length is of the order $\sim$tens kpc, i.e., of the order of the stellar disk size in spirals. Thus, a~quantum pressure emerges (\citep{deVega_2010,Destri_2013, de_Vega_2013, de_Vega_2016, Lovell_2014,destri2, de_Vega_2017}) that shapes the inner DM density profile forming a cored distribution that \citep{devega2012, Destri_2013} have well reproduced with the {\it pseudo-isothermal} profile: 
\begin{eqnarray}
\rho_{PISO} (r)= \rho_0 \frac{r_0^2}{(r^2+r_0^2)}
\label{PISO}
\end{eqnarray}
where $\rho_0$ is the central constant density and $r_0$ is the core radius\endnote{It is interesting to notice that, before then, the NFW profile emerged from simulations, the PISO profile was the favourite in modelling the DM halos around galaxies}. The rotation curves of the whole family of normal spirals are well reproduced by the above scenario with a particle mass of $\sim$2 keV \citep{de_Vega_2014}, see Figure \ref{WDM_profile}. Important lower limits for the mass of such fermionic particle are recently obtained by investigating the smallest dwarf spheroidal (dSph) satellites of the Milky Way and of Andromeda \citep{DiPaolo_2018_WDM,Alvey_2021}.

Finally, we have to report that there are claims of indirect detection for a 7 keV fermionic particle (e.g., \citep{Boyarsky_2014}).

\vspace{-2mm}

\section{WIMPS as DM Candidates?}

\label{Experiments}
 The structural properties of the dark matter in galaxies have recently become crucial to tackle the dark matter phenomenon in that the well known successes of the $\Lambda \rm CDM$ scenario on theoretical, numerical simulations, and cosmological sides have been negatively balanced by the outcome of carefully designed experiments and astrophysical observations aimed to detect the related WIMP particle. In fact, it is fair to state that none of them has yet succeeded (e.g., \citep{Arcadi_2018}). In the following we will give a brief account of this. 
There are three main possible ways to ``detect'' the DM particles:

\begin{itemize}
\item[({\em i})] {\em Indirect {Detection}}
\end{itemize}

This refers to the annihilation or the decay products of DM particles occurring far away from Earth in some DM halo including that of our own Galaxy. These efforts focus on locations where the DM is thought to accumulate the most, since the signal scales as $\rho_{\rm DM}^2$ for annihilations and as $\rho_{\rm DM}$ for decays: i.e., the centers of galaxies and clusters, as well as those of the smallest satellite galaxies of the Milky Way. Typical indirect searches look for excess of gamma rays, which are predicted both as final-state products of particles annihilation, or are produced when charged particles interact with ambient radiation via inverse Compton scattering. 
The spectrum and intensity of a gamma ray signal depends on the annihilation products and is computed on a model-by-model basis. 
The $\gamma$-ray flux of energy $E$ coming from dark matter annihilation in a distant source, extended within a solid angle $\Delta \Omega$ is given by
$$
\Phi_\gamma(E, \Delta \Omega) \propto [(\langle \sigma v \rangle / m_{_{WIMP}}^2 ) \sum_f b_f dN_{\gamma}/dE] \ J_{\Delta \Omega}
$$
where $\langle \sigma v \rangle$ is the thermally averaged annihilation cross-section and $b_f$ and $dN_{\gamma}/dE$ denote the branching fraction of the annihilation into the final state $f$ and the number of photons per radiated energy, respectively. In addition to the physical processes and the DM particle mass, the $\gamma$-ray flux also depends on the spatial DM distribution through the $J$-factor:
$$
J_{\Delta \Omega}= \int_{\Delta \Omega} \int_{los} dl \; \Delta \Omega \; \rho^2 (l, \Omega)
$$
 in case of a decay process the J-factor \citep{Gunn_1978, BERGSTROM_1998, Geringer-Sameth_2015}
 $$D_{\Delta \Omega}= \int_{\Delta \Omega} \int_{los} dl \; \Delta \Omega \; \rho (l, \Omega). $$
 
These factors correspond to the line-of-sight ($los$) integrated squared or proportional dark matter density within a solid angle $\Delta \Omega$.
Experiments have placed bounds on the DM annihilation or decay, via the non-observation of the annihilation or decay signals.
For constraints on the cross-sections, see {Figure} 2 in \citep{Hoof_2018} (Fermi-LAT), {Figure} 8 in \citep{Archambault_2017} (VERITAS), {Figure} 1 in \citep{Abdallah_2016} (H.E.S.S.), {Figure} 5 in \citep{Cui_2018} (AMS-02) and {Figure} 4 in \citep{Iovine_2019} (IceCube and ANTARES).

\begin{itemize}
\item[({\em ii})] {\em Direct Detection}
\end{itemize}

This refers to the effects of a DM particle--nucleus collision as the dark particle passes through a detector in an (underground) Earth laboratory. 
The WIMP elastically scatters off the atomic nucleus and the momentum transfer gives rise to a detectable nuclear \mbox{recoil \citep{Goodman_1985, Schumann_2019}.} Currently, there are no confirmed detections of dark particles from direct detection experiments (e.g., XENON1T, CDMSlite, COUPP, PICO60(C$_3$F$_8$), PICASSO, PANDAX-II, SuperCDMS, CDEX, KIMS, CRESST-II, PICO60(CF$_3$I), DS50, COSINUS, DarkSide-50), but only upper limits on the DM particle---standard model particle cross-section as function of the particle mass, see {Figures} 12 and 13 in \citep{Schumann_2019} and {Figure} 1 in~\citep{KANG_2019}.

\begin{itemize}
\item[({\em iii})] {\em Collider Production}
\end{itemize}

This approach attempts to produce DM particles in a laboratory. Experiments at the large hadron collider (LHC) could create them via collisions of the LHC proton beam. In this case, the DM particle would be detected indirectly as (large amounts of) missing energy and momentum escaping the detectors \citep{Kane_2008}. The resulting LHC and LEP constraints on the DM particle cross sections can be found in Figure 3 in \citep{Trevisani_2018} and in (e.g., \citep{Fox_2011}). 
 
In short, so far, we have not a WIMP-like dark particle detection but very careful upper limits on their cross section as function of the particle mass that exclude, as the dark particle, the most expected WIMP candidates. It is also fair to notice that there is a still large, though not theoretically favoured, WIMP range in (cross section, particle mass) not yet investigated.

\subsection{Observational Issues with WIMP Scenario }
\label{Issues_with_the_main_DM_scenario}

 The N-body simulations in the $\Lambda$CDM scenario produce results well in agreement with the large scale structure of the Universe (i.e., at scales $\gtrsim$1 Mpc),  however, they also predict an overabundance of small structures which may be not observed in dedicated surveys. This is the {\bf {missing} satellite problem} (e.g., \citep{Klypin_1999, Moore_1999, Zavala_2009,Papastergis_2011,Bullock_2010, Klypin_2015}). A possible explanation for this discrepancy is the existence of dark dwarf satellites that failed to accrete gas to form stars either because of the expulsion of the former in the supernovae-driven winds or because of gas heating by the intergalactic ionising background. 
 However, more massive halos with $M_{vir} > 10^{10} M_\odot$ have deep potential wells and should be able to retain the primordial gas and form stars. Nevertheless, also in this case we do not observe the large number of objects predicted by the N-body simulations. In short, the predicted luminosity function of sub-halos is not in agreement with observations. This is the {\bf too big to fail problem} (e.g., \citep{Ferrero_2012, Boylan_2012, Garrison_Kimmel_2014, Papastergis_2015}). 

Furthermore, the inner dark halo density cusps predicted from the N-body simulations are in strong contrast with the observed cored density profiles, well described by the Burkert law (Equation (\ref{V_dark_matter1})) 
This is the well-known {\bf cusp-core problem} (e.g., \citep{Salucci_2001, deBlok_2002, Gentile_2004, Gentile_2005,Simon_2005, Del_Popolo_2009, Oh_2011,Weinberg_2013}, for a recent review \citep{Salucci_2019}) present in spirals of any luminosity (\citep{niloufar_2020}). The solution of this issue, proposed within the CDM-WIMP scenario, involves the late-time effect of {\bf baryonic feedbacks} on the primordial cuspy DM distribution: these are generated by supernovae explosions which blow the existing gas to the outer galactic regions, rapidly modifying the total gravitational potential and, in turn, erasing the inner cusp of the dark halo density \mbox{(e.g., \citep{Navarrot_1996, Read_2005, Mashchenko_2006, Pontzen_2014, Di_Cintio_2014}).}
Let us stress, however, that this process is unable to produce the observed cored DM distribution in dwarf and large spirals \citep{ Di_Cintio_2014}. Furthermore, the halo response to the stellar feedback is shown to be a strong function of the star formation threshold~\citep{Dutton_2019, Benitez_2019} and this rises doubts on its ability to form a cored dark halo distribution in any galaxy. On this perspective, the DM distribution in LSB galaxies, with a very low star formation rate, is a crucial test for an astrophysical solution of the core-cusp issue.

\subsection{Issues with NO-WIMP Dark Particle Candidates}
\label{Issues with other DM candidates}

The DM reference particle is cold and collisionless, however, it is interesting to note that also alternative scenarios to $\Lambda \rm CDM$ run in difficulties and this makes the information on the DM particle that we can extract from the LSBs structural properties even more important, by providing us with additional clarifying tests for the various particle~candidates.

The ULA scenario is challenged by the existence of DM core radii with size\linebreak $\gtrsim$10 kpc~\citep{Hui_2017,Burkertula}. The SIDM is strongly constrained by clusters observations \citep{Banerjee_2019}. At galactic scales it requires a fine-tuned velocity dependence of the self-interacting cross section, without which, the cores of the DM halos would have all the same size, determined by the elementary particle physics. 

 Challenges to the WDM scenario emerge at the level of the big bang nucleosysthesis~(\citep{Sabti_2020}) and at intermediate redshifts (e.g., \citep{Irsic_2017,Enzi_2020}), although the very quantum nature of this particle has to be fully investigated yet.

\section{The Dark and the Luminous Matter Distribution in Disc/LSB Galaxies}
\label{Galaxies_the_luminous_and_dark_matter_distribution}

\begin{figure}[H]
\smallskip
\includegraphics[width=0.66\textwidth,angle=0,clip=true]{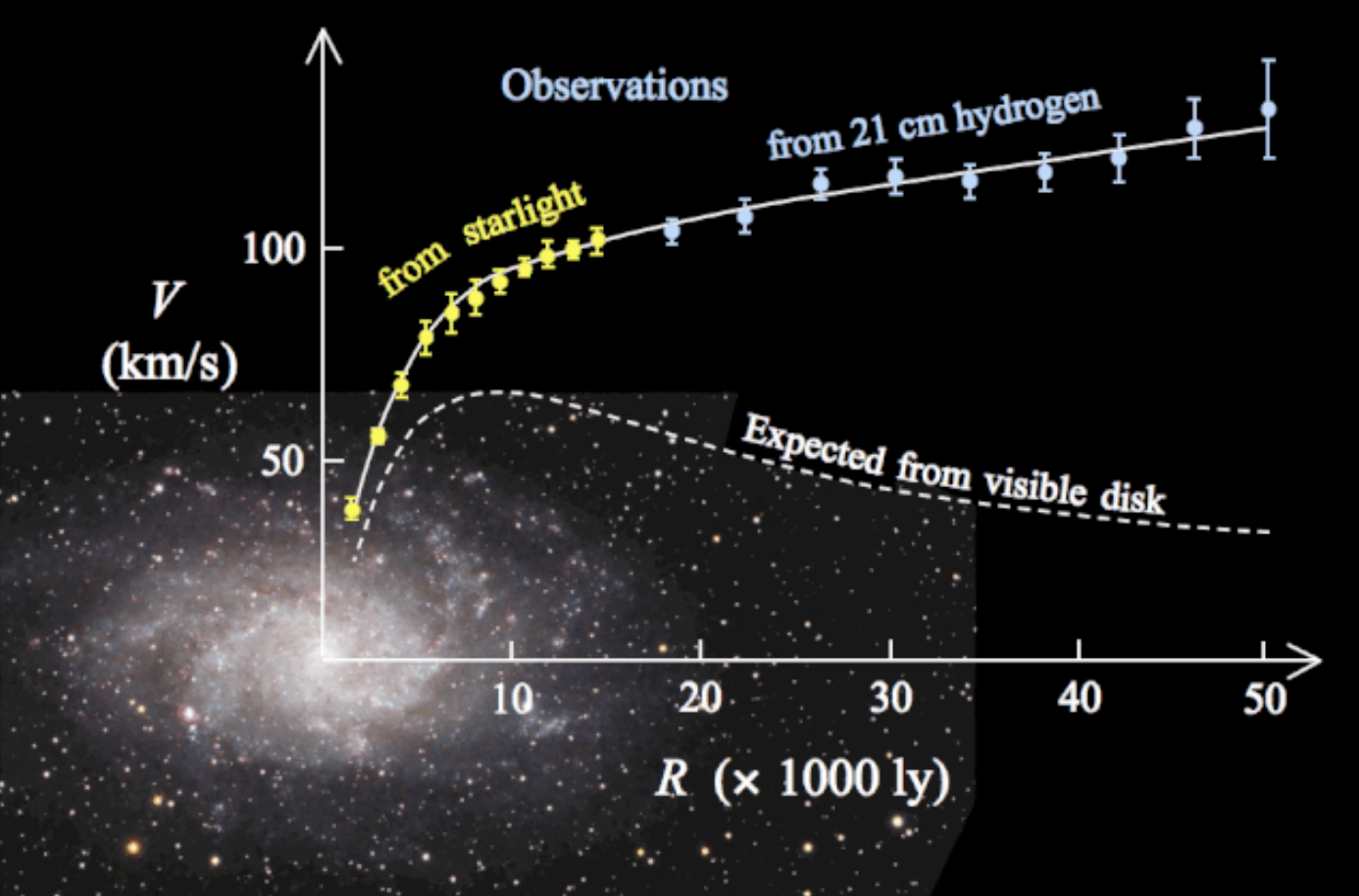}
\caption{Rotation curve (km/s vs. light years) of the Spiral M33 obtained from optical and 21 cm data (points with error bars) compared with the RC expected from the luminous component (dashed line). The solid line is the RC model when a dark halo is {included} (see \citep{Corbelli_2000} for details).}
\label{SMC_rot_curv_fig2}

\end{figure}

One important way to investigate the DM properties is to study its distribution in galaxies. This is relatively direct in rotational supported systems, such as spiral galaxies, due to their rather simple kinematics (see Figure \ref{SMC_rot_curv_fig2}). In these objects one can obtain the dark and luminous matter distribution by best-fitting their rotation curves $V(R)$ with suitable models.\endnote{In this review we consider the RC and the circular velocity as equivalent quantities, assumption not allowed in other contexts.}  The circular velocity is directly related to the total galaxy gravitational potential~$\Phi(R)$ by:
$$V^2(R)=R \, d\Phi (R)/dR$$

This is different from the situation in elliptical or dwarf spheroidals whose kinematics or dynamics are dominated by random motions rather than by ordered rotational ones so that the determination of the mass distribution involves the velocity dispersion $\sigma(R)$ rather than the circular velocity $V(R)$.
From the Jeans equation we get the relation between the circular velocity $V(R)$ and the radial velocity dispersion $\sigma_r(R)$\endnote{We neglect here for simplicity the projection effects.} measured by a gravitational tracer with spatial distribution
$\nu_\star(R)$ and kinematical anisotropy $\beta(R)$:
 $$R \ \frac {d\Phi (R)}{dR}=V^2(R)=(-\gamma_\star (R) + 2\, \beta(R)+2 \, \alpha(R))~ \sigma_r^2(R)$$ 
 where $\alpha(R)$ and $\gamma_\star(R)$ are the logarithmic derivatives of $\sigma_r$ and $\nu_\star$. We realise that in this case the study of the kinematics is complex and complicated by the presence of the unknown anisotropy parameter, all this causing strong degeneracies in the mass modelling (e.g., \citep{Binney_1976, Binney_1978, Kormendy_1996, Cappellari_2007}).

 In disk systems, with $ i=(d,HI,bu,h)$ where $V_d$, $V_{HI}$ $V_{bu}$ and $V _h$ being the contribution in quadrature to the total circular velocity $V(R)$ due to the stellar disc, the gaseous disc, the bulge, and the dark matter halo and with $b$ indicating the total baryonic contribution, we have: 
\begin{eqnarray}
\label{Spirals_form4}
V^2(R)=R \sum^i\frac{d\Phi _i}{d\,R}= V^2_{mod}=V^2 _d (R)+ V^2 _{HI} (R)+ V^2 _{bu} (R)+ V^2 _h (R)=V^2_b(R) +V^2_h(R)
\end{eqnarray}

The Poisson equation: 
$$\nabla^2 \Phi_i=4 \pi \ G \rho_i$$ 
relates, component by component, the surface and volume densities to the corresponding gravitational potentials 
 (\citep{Salucci_2019}). In the process of determining the galaxy mass model we obtain, directly from the galaxy photometry, the {\it radial profile} of the ``luminous components'' (i.e., $V_d(R)^2/M_D$ and $V_{bu}(R)^2/M_{bu}$) with the disk and the bulge masses as free parameters of the fitting model. The measured HI surface density, in objects with well known distance, directly yields $V _{HI}(R)$. 
 
Let us stress again that the observed rotation curve of a disk system like a LSB galaxy, firstly, is assumed to well represent the galaxy circular velocity $V(R)$ and then is fitted by the model velocity curve $V_{\rm mod}(R)$ of Equation (\ref{Spirals_form4}) that includes the various baryonic contributions alongside with that of the dark halo (all in quadrature). 
 
It is also useful to remark that, in rotating systems, the galaxy total gravitational potential $\Phi(R)$ relates 
with the radial acceleration $g(R)$ of a point mass at distance $R$ and with its baryonic contribution $g_b(R)$ according to
\begin{eqnarray}
g(R) = V^2(R)/R = | - d \, \Phi(R) / d \, R |,
g_b(R) = V^2_b (R) / R = |-d \, \Phi_{b}(R) / d \, R |
\label{DM_gravity}
\end{eqnarray}

\subsection{The Stellar Disc}
\label{The_stellar_disc}

Given a galaxy stellar disc with a known surface density profile its contribution to the circular velocity is obtained from the Poisson's equation in cylindrical coordinates (see Equation (3) of \citep{Kent_1986}).
Caveat some occasional cases, not relevant for the present topic, the stars in rotating systems are mainly distributed in a thin disc with surface luminosity following, in a specific X Band (e.g., the B-Band), the Freeman profile \citep{Freeman_1970}:
\begin{eqnarray}
\label{Spirals_form1}
\mu(R)= \mu_{0,X} \, e^{-R/R_D}
\end{eqnarray}
and $R_D$ is the disc scale length (for LSBs, see {Figures} 1 and 2 in \citep{McGaugh_1994b} and {Figures} 7--11 in~\citep{Wyder_2009}). The length scale $R_D$ does not depend on the band X, especially if this is in the IR region of the spectrum. 
The disk surface density is then given by: 
$$(M_D/L)_X \ \mu(R)$$
with the first term is the mass-to-light in the X band and $M_D=2 \pi R_D^2 \mu_{0,X} (M/L)_X$ is the disk mass. Therefore, expressed in the radially normalised units $r/R_D$, the Freeman light {\it profile} does not depend on the galaxy luminosity; in all objects the disc length $R_D$ sets a consistent reference scale for the stellar disk distribution. It is useful to define the optical radius $R_{\rm opt}\equiv 3.2 \ R_D$ as the stellar disc size: this radius encloses, in any object, 83\% of the total disc galaxy luminosity. Notice that $R_{opt}$ and the often used (in spheroidal galaxies) half-light radius $R_{1/2}$ enclosing half of the total galaxy luminosity, are both good tags of the luminous size of an object; for Freeman disks: $R_{1/2}=1.68 \, R_D$.
The contribution to the circular velocity from the stellar disc component is given by:
\begin{eqnarray}
\label{V_stellar}
V_d^2(R) = \frac{1}{2} \frac{G\, M_D}{R_D} (3.2\ x )^2 (I_0 K_0 - I_1 K_1),
\end{eqnarray} 
where $I_n$ and $K_n$ are the modified Bessel functions computed at $1.6 \,x$, with $x = r/R_{opt}$.

\subsection{The Gaseous Disc}
\label{The_gaseous_disc}
\begin{figure}[H]
\smallskip
\includegraphics[width=0.72\textwidth,angle=0,clip=true]{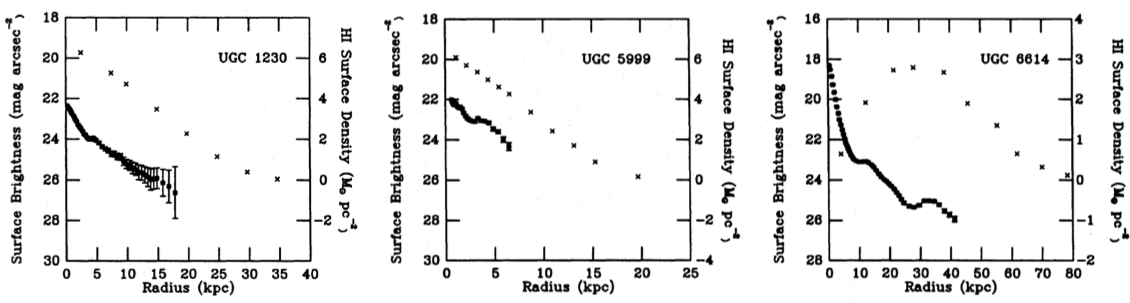}
\caption{The radial distribution of the R-band surface brightness and the HI surface density in three LSBs. The image is reproduced {from} \citep{vanderHulst_1993} (Figure {2}).} 
\label{fig6}
\end{figure}

A gaseous HI disc is present in rotating disc galaxies. $V_{HI}$, the contribution to the circular velocity is obtained from the HI surface density $\Sigma_{HI}(R)$ by solving, as for the stellar disk, the corresponding Poisson's equation \citep{Kent_1986}. 
Typical gas distributions are shown in Figure \ref{PWS}. Very approximately, in the external regions, the gaseous HI disc shows a Freeman profile (see {Figure 2} of \citep{vanderHulst_1993}) with a scale length about three times larger than that of the stellar disc in the same object \citep{Evoli_2011, Wang_2014}, so that: $\Sigma_{HI}(R)= \Sigma_{HI,0}e^{-R/3\,R_D}$. In this case the contribution of the gaseous disc to the circular velocity is: 
$$V^2_{HI}(R)=1.3 \frac{1}{2} \frac{G M_{HI}}{3 R_D} ( 1.1 \, x)^2 (I_{0} K_{0}-I_{1} K_{1})$$
where the factor 1.3 takes in consideration the helium contribution to the gaseous galaxy disc, $M_{HI}$ is the HI disc mass, $I_n$ and $K_n$ are the modified Bessel functions computed at $0.53 \, x$.
This assumption is especially valid for for objects with $V_{opt} \gtrsim 150 \,$ km/s in that in them this component is quite secondary: their star formation has been very efficient in turning the primordial HI disk in a stellar one. 
In the very outer regions ($x>2$), in the circular velocity of any disk system, including the LSBs, the gas component overcomes the stellar one, although at these radii, both contributions are negligible with respect to the component from the DM halo \citep{Evoli_2011}. 
Furthermore, especially in LSBs, the HI disc is important as tracer of the galaxy gravitational field because of its extension into regions where the $H_\alpha$ kinematical measurements lack (see Figure \ref{fig6}). 
Finally, inner H$_2$ and CO discs are also present and in some case might be of some relevance with respect to the stellar and HI ones (\citep{Gratier_2010}).

\subsection{The Stellar Bulge}
\label{The_stellar_bulge}
Large disc galaxies are characterised by the presence of a central bulge, which usually appears as a round ellipsoid, where old and new stars are crammed tightly together within few hundredths of parsecs. The mass profile decomposition must take in consideration that we have an very inner projected stellar density $(M/L)_{X} \mu_{bu,X}(R)$ which gives a contribution to the circular velocity, as specified in Section 5 of \citep{Kent_1986}. Noticeably, far away from the center, $V^2_{bu}(R)= G \frac{M_{bu}}{R}$, where $M_{bu}$ is the bulge mass.
Assuming that the innermost velocity measurements are obtained at a radius $r_{in}$ is larger than the edge of the bulge, we can consider it as a point mass, especially in LSBs where the stellar/HI disks are very spatially extended. The contribution $V_{bu}$ to the circular velocity, relevant in the inner galactic regions of the biggest LSBs, is then:
\begin{eqnarray}
\label{Bulge_velocity}
V^2_{bu}(R)= \alpha _{bu} V^2_{in} \left(\frac{R}{r_{in}} \right)^{-1},
\end{eqnarray} 
where $\alpha_{b}$ is a parameter which can vary from $0.2$ to $1$ (e.g., see \citep{Yegorova_2007}), $\;V_{in}$ and $\;r_{in}$ indicate the values of the velocity measurement closest to the galactic center. 

After we caution that, to model the bulge with 3 free parameters without assuming very strict priors on them may cause a large degeneracy in the resulting galaxy best fit model we introduce the Sersic bulge 
$$\Sigma_{bu} (R)=\ I_{0,bu} Exp (-(2 n-1/3)(R/R_{1/2})^{1/n})$$ one has \citep{Nord}:
\begin{eqnarray}
\label{eqn:5}
 V_{bu}^2(R)=B\int_{m=0}^{R}\left[ \int_{\kappa=m}^{\infty}\frac{e^{-(\kappa/R_{1/2})^{1/n}}(\kappa/R_{1/2})^{1/n-1}}{\sqrt{\kappa^2-m^2}}d\kappa \right]\frac{m^2}{\sqrt{r^2-m^2e^2}}dm
\end{eqnarray} 
where $R_{1/2}$ is the projected radius enclosing 50\% of the bulge light, $B=\frac{4\ G \ (M/L)_X I_{0,bu,X}}{R_{1/2} \,n} $.

\subsection{The DM Halo}
\label{The_DM_halo}
Since the luminous component is not able to fit the whole rotation curve (\citep{Rubin_1980, Bosma_1981,Salucci_2019}) we need to add a contribution by an assumed {\it spherical dark matter halo}.
The contribution to the total circular velocity is given by\endnote{Since in spirals the kinematics is all in the rotation plane, the spherical coordinate $r$ coincides with the cylindrical coordinate $R$}:
\begin{eqnarray}
\label{V_dark_matter}
M_{DM} (r) = \int_0^r 4\pi {\tilde r}^2 \rho _{DM} ({\tilde r}) \; d {\tilde r}
\\
\label{V_dark_matter1}
V_h^2(r) = G \frac{M_{DM}(r)}{r} \quad.
\end{eqnarray} 
with $M_{DM} (r)$ the DM mass profile. The density profiles $\rho_{DM} (r)$ mostly used are: 
 
\begin{itemize}
\item[(i)]  {The} {\it NFW profile}, described by Equation (\ref{NFW_profile}) which is the popular fit of the outcome of N-body simulations in the $\Lambda$CDM scenario. It is characterised by a central {\it cusp} $\propto r^{-1}$ and by an external tail $\propto r^{-3}$; in more detail, we have that, in simulations, in the interval $2.5 \ r_s<r<R_{vir}$: 
$\rho_{\rm NFW}(r)\propto r^{-2.7^{+0.1}_{-0.1}}$
where the upper and lower limits $\pm 0.1$ originate from the different values, among halos, of the concentrations $c$ (see below) and, in each halo, of the radius $r$;
 
\item[(ii)] {\it empirical cored profiles} characterised by a central constant density within a core radius $r_0$ (i.e., $\rho(r) \simeq const$ for$ \ r< r_0$) and by an external tail whose negative slope can vary according to the specific adopted model. 
\end{itemize}

The pseudo-isothermal profile (see Equation (\ref{PISO})) and the Binney--Tremaine profile \mbox{(see \citep{Binney_2008,Persic_1996})} are often used,
however, these profiles, despite that successfully help fitting the kinematics inside $\sim$2 $r_0$, are characterised by an external {tail} $\propto r^{-2}$ (that implies constant circular velocities when $r\gg R_{opt}$) and then disagree with the declining RC profiles found at outer radii and that are also supported by weak lensing and other astrophysical measurements \citep{Hoekstra_2008, Zu_2015, Donato_2009,Shankar_2006,Salucci_2007, Zobnina_2019}. Finally, they also do not comply with the decline in the outermost regions shown by NFW profile. 
 
A very successful empirical model is the {\it Burkert profile} \citep{Burkert_1995,Salucci_2000,Salucci_20000} {that}, at large scales, converges to the (collisionless) NFW one: 
\begin{eqnarray}
\rho _{B} (r) = \frac{\rho _0 r_0 ^3 }{(r+r_0)(r^2+r_0^2 )} 
\label{DM_density}
\end{eqnarray} 
where $\rho_0$ is the central mass density and $r_0$ is the core radius. Additionally, this profile is characterised by an external tail $\propto r^{-2.7}$. The corresponding mass profile is:
\begin{eqnarray}
\label{DM_Mass}
M_{B} (r) = 
2 \pi \rho_0 r_0 ^3 \; [ln (1+r/r_0 ) - tg ^{-1} ( r/r_0 ) + 0.5\; ln (1+( r/r_0) ^2 )]. 
\end{eqnarray}

The Burkert profile well represents the family of the cored halo distributions: noticeably, {\it inside} $r_0$, cannot be discriminated from the other cored profiles including the ``theoretical'' ones occurring in the cases of degenerate fermionic particles or boson condensates (see Appendices A1 and A2 in \citep{DiPaolo_2018_WDM}), \citep{de_Vega_2013,de_Vega_2014,de_Vega_2017,Schive_2014}. Of course, despite that the circular velocity fits can be very similar independently of the assumed (cored) profile, the resulting 3D relationship: central density-core radius-halo mass is instead very density profile dependent. 
Outside $r_0$ the Burkert profile converges to the NFW one, this could be explained by the fact that, in the external regions of galaxies, the distances among particles are so large that the DM halos are, on a Hubble time, collisionless also if the individual particle are not.
Remarkably, the Burkert profile well reproduces, in cooperation with the velocity components of the luminous matter, the {\it individual} and {\it stacked } circular velocities of spirals, dwarf disks, and of a number of ellipticals (see \citep{Karukes_2017,Salucci_2018,Memola_2011}).

It is worth to briefly discuss the {\it Zhao profile} \citep{Zhao_1996}: $$\rho_{Z} (r)= \frac{\rho_0}{\big( r/r_0 \big)^{\gamma} \left(1+ (r/r_0)^{\alpha} \right)^{\frac{\beta + \gamma}{\alpha}}}$$ 
that is claimed to reproduce the NFW, the Burkert, and other cored or cuspy profiles just by tuning the values of its free parameters and to be very apt to cope with different levels of cuspiness present in the densities of the dark halos. However, this is obtained at the expenses of severe malfunctions. First, this profile involves a large number of parameters: the ``central'' density $\rho_0$, the core radius $r_0$ and $\alpha$, $\beta$, and $\gamma$, three parameters that control the slope and the curvature of the profile. Such large number of parameters is in disagreement with the circular velocity data that always are well fitted with the help of DM halo profiles with just 2, rather than 5, free parameters. Moreover, simulations and investigations in $\Lambda \rm CDM$, FDM and WDM scenarios (e.g., \citep{Navarro_1997, Schive_2014, de_Vega_2014}) show that the resulting DM density halo profiles are represented by a general function of radius with, not more than, two parameters that run differently in each halo. On a numerical case, such over-fitting, coupled with the observational errors in the kinematical data, causes strong degeneracies in the values of the best fit parameters.

A most relevant dark halo quantity is its mass. Cosmologists refer to the {\it virial mass} $M_{vir}$ that evaluated from the halo mass profile $M_{DM}(r)$ according to the relation (e.g., \citep{Mwhite}): $M_{\rm DM}(R_{vir})= \frac{4}{3} \, \pi \,200 \, \, \rho_{c} \, R_{vir}^3$, where $R_{vir}$ is the virial radius and \mbox{$\rho_{c} = 9.3 \times 10^{-30}$ g/cm$^3$} is the critical density of the Universe assumed here.

\subsection{RC Analysis}
\label{Rotation_curves_modedeling}

The stellar disk mass (or the quantity $(M_\star/L)_X$) is one of the 3 free parameters of the {\it fitting mass model} that we use to reproduce the various RCs. We notice that $M_\star$ can be inferred also in other ways: 

\begin{itemize}
\item[(i)] From galaxy colours (or spectral energy distributions) by means of the stellar population synthesis models (e.g., \citep{Drory_2004,Salucci_2008};

\item[(ii)] From the maximum disc hypothesis, according to which, inside $2 R_D$, the stellar disk takes the maximum possible value $M_{D,max}$, under the constraint that, at any radius, $ V_d \leq V(R)$ (see \citep{PS90}).
\end{itemize}
 
 In both cases the number of the required fitting parameters gets reduced by one. The two free parameters adopted in the DM halo mass model are obtained as result of the best fitting of {\it individual} or {\it coadded}\endnote{For some author coadded = stacked} RCs. Noticeably, in disk systems, the values of the structural parameters of the LM and DM mass distributions, obtained by modelling either (i) an ensemble of coadded RCs or (ii) a large and proper sample of individual RCs, well agree. Moreover, their combined knowledge enlighten the properties of the galaxies' mass~components.

\section{The Universal Rotation Curve of LSB Galaxies}

\label{Universal_rotation_curve}

\begin{figure}[H]
\smallskip
\includegraphics[width=0.65\textwidth,angle=0,clip=true]{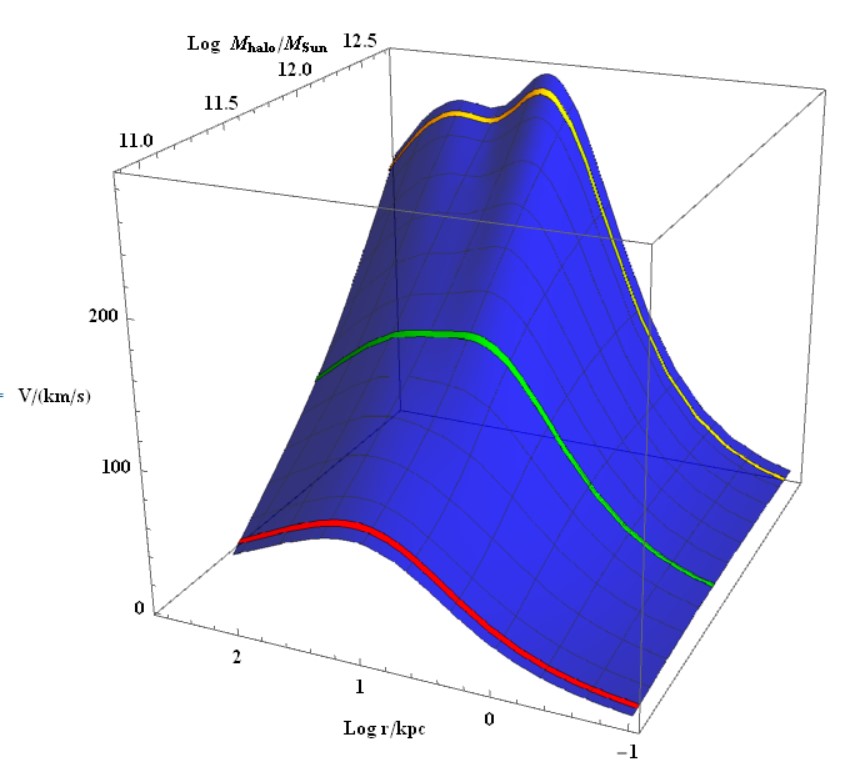}
\caption{{The universal} rotation curve (URC) for spirals. The circular velocity $V(r; M_{vir})$ is plotted (\textbf{right} to \textbf{left}) out to $R_{vir}$ as a function of log radius and $Log M_{vir}$. The 3 RCs plotted as yellow bands correspond to the cases $Log\, M_{vir}/M_\odot =11 \pm 0.03, 11.9 \pm 0.03,12.3\pm 0.03 $. }
\label{RC_Salucci}
\end{figure}

It is well known that disc RCs of galaxies of different magnitude $mag$ ({\bf or} different velocity $V_{opt}$), especially when expressed with their radial coordinate $R$ normalised to their optical radii $R_{opt}$, follow an {\it Universal} trend (first shown in {Figure} 4 of \citep{Rubin_1985}, then in~\citep{Persic_1991,Salucci_2019,Persic_1996, Roscoe_1999, Catinella_2006, Noordermeer_2007, Salucci_2007, Lopez_2018, Karukes_2017,Lapi_2018},  and Figure \ref{RC_Salucci}). The universal rotation curve is the analytical curve that catches such a trend can be expressed both in physical units: $$ V_{URC}= F(R,[M_{vir}\, {\bf or} \, mag\, {\bf or} \, V_{opt}])$$ and in normalised units: $$ \frac {V_{URC}}{V_{opt}}=F_N\Big(\frac {R}{R_{opt}},[M_{vir}\, {\bf or} \; mag\; {\bf or} \; V_{opt}]\Big)$$
 
\vspace{-18pt}

It is worth to underline that {\it in relation to the issue of the dark matter phenomenon} the concept of {\it universality} of the RCs means that every RC\endnote{That can be both an individual RC of an object with ($V_{opt}$, $R_{opt}$) that we indicate with: $V_{ind}(R; R_{opt}, V_{opt})$, or the RC emerging from the coaddition of many RCs of objects with similar optical velocities and optical radii (whose averaged values are (<$V_{\rm opt}$>, <$R_{\rm opt}$>)) that we indicate with: $V_{coadd}(R; <R_{\rm opt}>,< V_{\rm opt}>)$
} can be described by a general analytical function of (a) the normalised radius $R/R_{opt}$ and of (b) a second observational quantity (e.g., $ V_{\rm opt}$). These two quantities tag perfectly the systematics of the velocity and mass distributions of the whole family of disk galaxies, so that, for any individual or coadded RC data we have: $V_{ind,coadd}(R;R_{\rm opt}, V_{\rm opt})\simeq V_{\rm URC}(R; R_{\rm opt}, V_{\rm opt})$. Such successful modelling of the kinematics of disk galaxies allows us to derive their mass distributions and, in turn, the properties of their DM halos by means of the {\it ``Universal Rotation Curve} (URC) method''. The latter involves two steps: (a) the coaddition process, i.e., the grouping of similar and properly normalised RCs and (b) the subsequent mass modelling of these coadded curves that takes advantage of the fact that we adopt for $F_N$ the sum in quadrature of the baryonic and dark matter velocity components of the circular velocity. 

The URC method has relevant advantages over the fitting of the {\it individual} RCs, if we are aimed to investigate the systematics of the luminous and dark matter distributions in disk galaxies. In fact, this statistical procedure increases the signal-to-noise ratio and smooths away the small-scale fluctuations induced in the RC by bad data and by physical features unrelated to the DM as spiral arms or disk warps. Moreover, one can include in the investigation RCs that cannot be fitted individually due to their scarcity of~measurements.

The URC is a very powerful tool to describe the distribution of mass in disk systems: in fact, once it is established for a family of galaxies, after we measure in an object $R_{\rm opt}$ and $V_{\rm opt}$, the tags that specify a galaxy, we can derive and predict its full rotation curve $V_{\rm URC}(r,R_{\rm opt},V_{\rm opt};\rho_0, r_0, M_D) $ in that the 3 structural dark and luminous parameters result all strongly correlated with $V_{\rm opt}$ and $R_{\rm opt}$. The URC method was applied for the first time in~\citep{Persic_1991}, this was followed by \citep{Persic_1996,Salucci_2007} that established the URC in (HSB) {\it normal spirals} see Figure \ref{RC_Salucci}. Previous results were confirmed and extended by \citep{Lapi_2018} with 2300~RCs of disc galaxies;~Ref. \citep{Karukes_2017} established the URC in 
{\it dwarf disc} ({\bf dd}) galaxies. It is worth stressing a particular result emerging in the above works: the discrepancies between the individual and coadded RCs and the corresponding ones predicted from the URC via their $V_{\rm opt}$ and $R_{\rm opt}$ values are about at a level of 7\%, a large part of which due to observational errors or to non-asymmetric motions present in the individual RCs. The URC results for LSBs \citep{DiPaolo_2019} will be discussed in the next sections.

\section{Low Surface Brightness (LSB) Galaxies}
\label{Low_Surface_Brightness_(LSB)_galaxies}
LSB galaxies are generally isolated systems, located at the edges of large-scale structure~\citep{Bothun_1997, Rosenbaum_2004, Galaz_2011, Kov_cs_2019}, near large-scale voids. 
During their formation in under dense regions, processes like tidal interactions, and mergers that increase the galaxy gas density rarely occur. Isolated environments characterise the giant LSBs \citep{Rosenbaum_2009}, while the LSB dwarfs and irregular galaxies are found in both under dense regions \citep{Pustilnik_2011} and more crowded environments \citep{Merritt_2014, Davies_2016}. This cosmological evidence implies that, in these systems, the primordial properties could have been conserved during the Hubble time, providing us with crucial information on the process of galaxy formation and structure evolution.

\begin{figure}[H]
\includegraphics[width=0.63\textwidth,angle=0,clip=true]{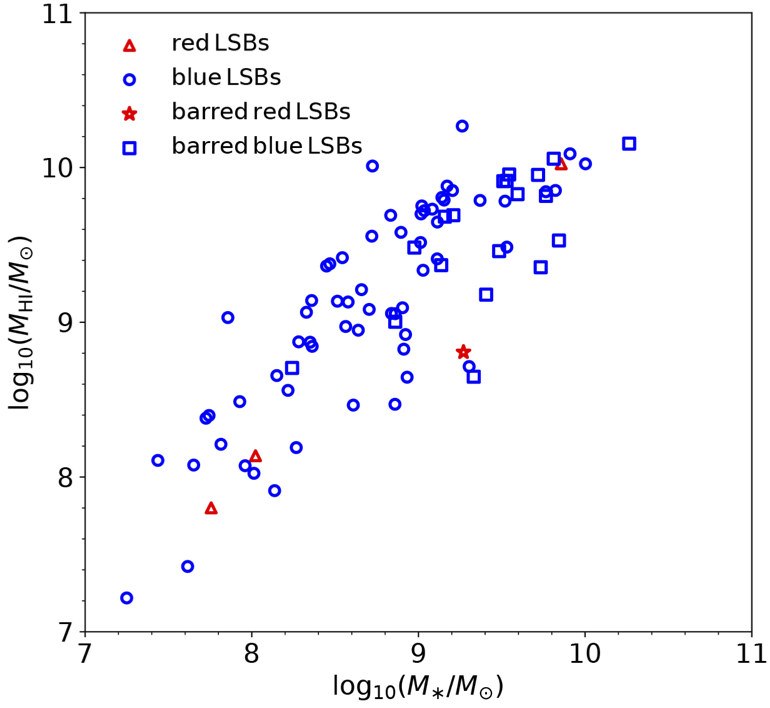}
\caption{The HI disk mass $M_{HI}$ vs. the stellar disk mass $M_\star$ for LSBs of different morphologies. Reproduced {from} \citep{Pahwa_2018}. } 
\label{MH_MD}
\end{figure}

\begin{figure}[H]
\includegraphics[width=0.6\textwidth]{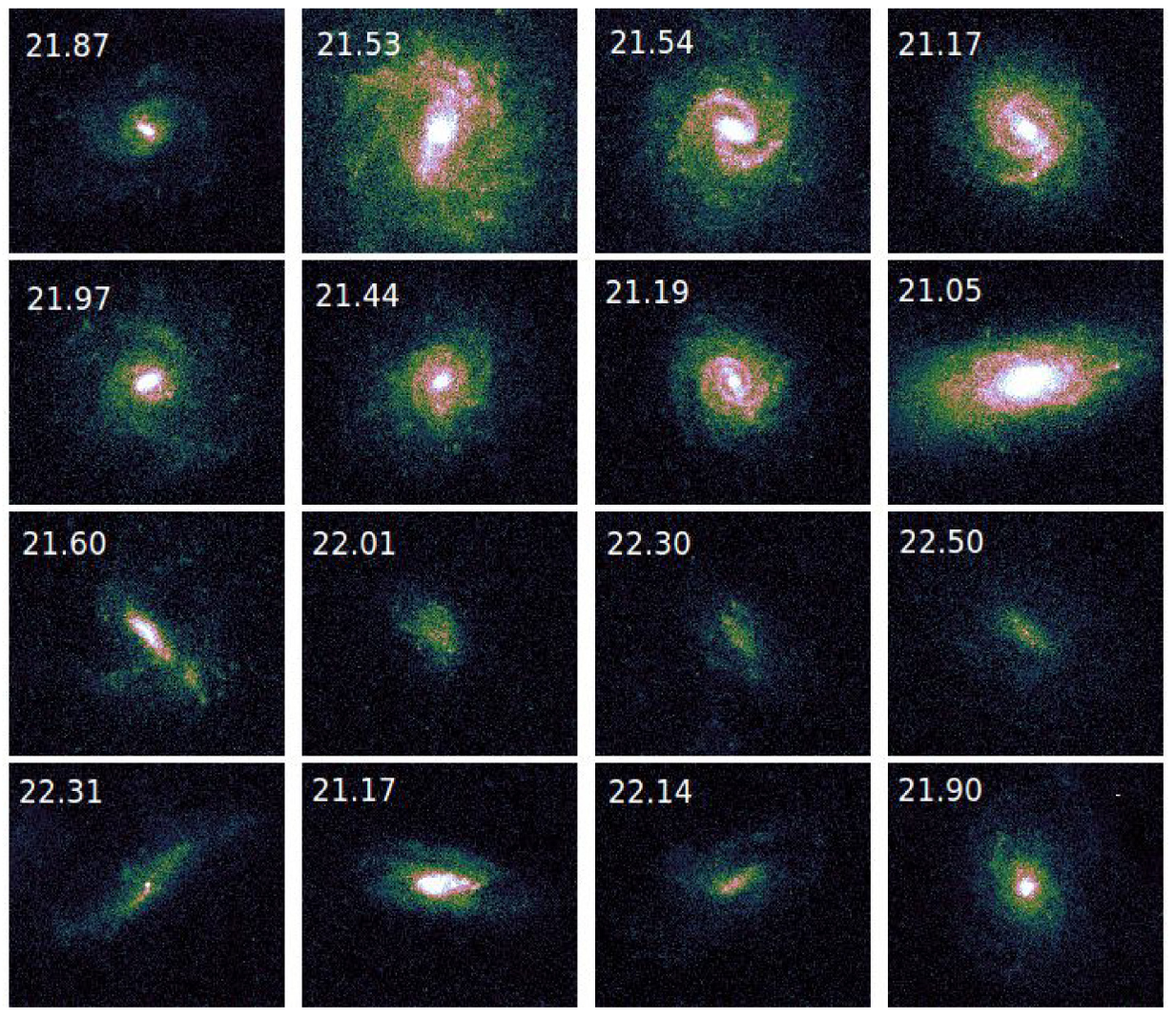}
\caption{R-band images of representative LSBs. The disc central surface brightness (in units of $\rm mag \, arcsec^{-2}$) is indicated. Image reproduced {from} \citep{Pahwa_2018}.} 
\label{Many_morphologies_LSB}
\end{figure}

\begin{figure}[H]
\includegraphics[width=0.65\textwidth]{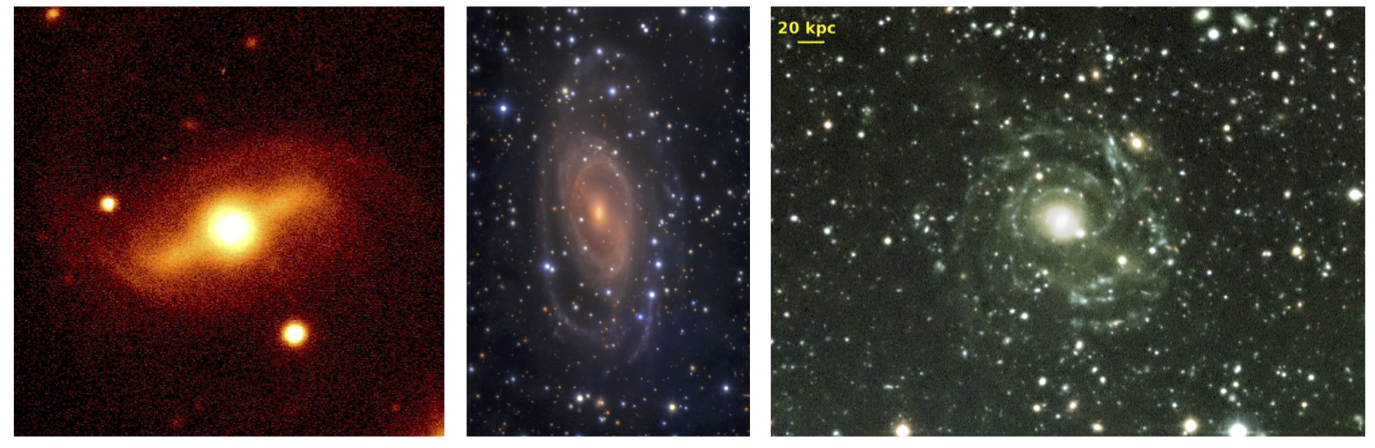}
\caption{{Peculiar} LSBs ({\bf left}) the barred LSB galaxy UM163 from \citep{Honey_2016}. ({\bf center}) UGC 1378 \citep{Saburova_2019} a giant galaxy with a central high surface brightness disc surrounded by an extended LSB disc. ({\bf right}) The giant LSB galaxy Malin 1 from \citep{Boissier_2016}.} 
\label{UGC1378}
\end{figure}

LSB galaxies (see {Figures} \ref{MH_MD}--\ref{UGC1378}) are rotating disc systems which emit an amount of light per area smaller than normal spirals, with a face-on central surface brightness $\gtrsim$23$\, \rm mag \,arcsec^{-2}$ in the B band (e.g., \citep{Impey_1997}) and $\rm \gtrsim$21$\rm \, mag \,arcsec^{-2}$ in the R band (see, e.g., {Figure} \ref{fig6} and also {Figures} in \citep{McGaugh_1994b,Wyder_2009}). In these objects the central surface brightness $\mu_{0,B}$ is significantly fainter up to 5 magnitudes down than the canonical value of \mbox{$\mu_{0,B} = 21.65\, \rm mag \,arcsec^{-2}$} of normal spirals \citep{Freeman_1970, vanderKruit_1987}. The oldest LSB galaxy samples were mainly composed of LSBs in the brightest end of surface brightness \mbox{(e.g., \citep{Schombert_1992, McGaugh_1994b, deBlock_1995, Impey_1996}).} Recently, LSB surveys comprise objects with much lower surface brightness ($\mu_{0,B}$ =\linebreak 24--28$\, \rm mag \,arcsec^{-2}$, e.g., \citep{Zhong_2008, Williams_2016, Trujillo_2016}). 

\vspace{-6pt}

\vspace{-4mm}

\begin{figure}[H]
\includegraphics[width=0.57\textwidth]{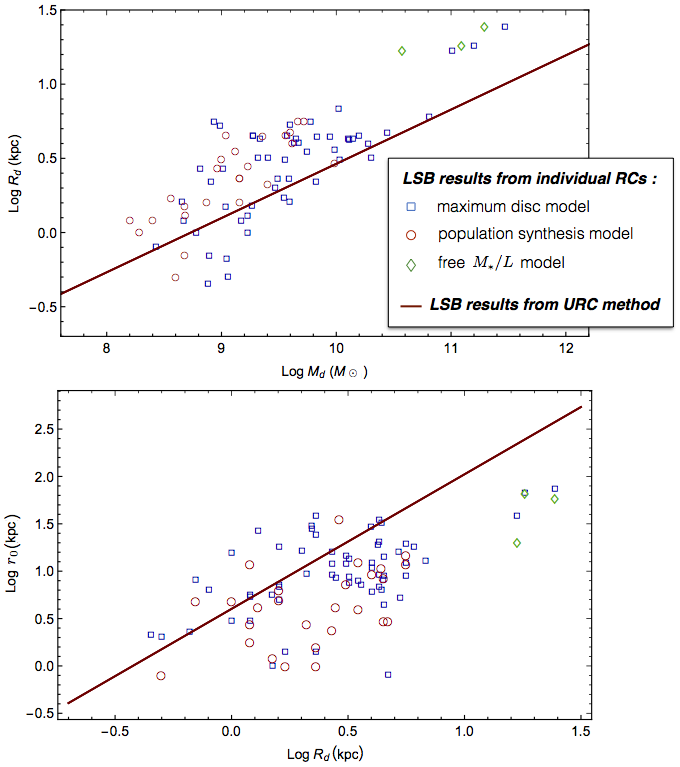}
\caption{Structural properties of LSBs obtained by best-fitting individual RCs. (\textbf{upper} panel): stellar disc scale length vs. mass of the stellar disc. (\textbf{lower} panel): halo core radii vs. the stellar disc scale lengths. Different symbols and the line play as in the inset \citep{DiPaolo_2019}. Image reproduced {from} \citep{DiPaolo_2020_PhD}.}
\label{Previous_work_Md_Rd}
\end{figure}

\begin{figure}[H]
\includegraphics[width=0.52\textwidth]{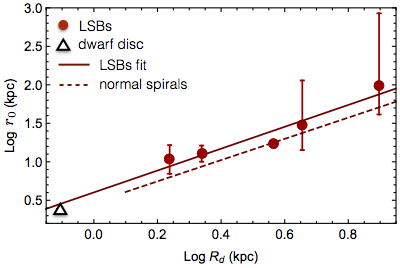}
\includegraphics[width=0.52\textwidth]{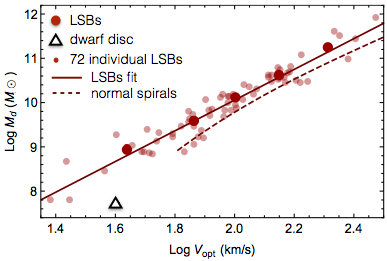}

\caption{(\textbf{up}): DM halo core radius vs. stellar disc scale length: data from the 5 coadded LSB RCs (points with errorbars) and their best fit (solid line) alongside with the same best fit in spirals (dashed line) and dwarf disks (triangle) \citep{Karukes_2017,Salucci_2018}). (\textbf{bottom}): stellar disc mass vs. the optical velocity. Legend as in the upper plot, also shown the individual data for the 72 LSBs of the sample (small circles). The spiral's best fit has a statistical uncertainty of $0.1 $ dex. Image reproduced {from} \citep{DiPaolo_2019}.}
\label{RC_RD_relation}
\end{figure}

\begin{figure}[H]
\includegraphics[width=0.55
\textwidth,angle=0,clip=true]{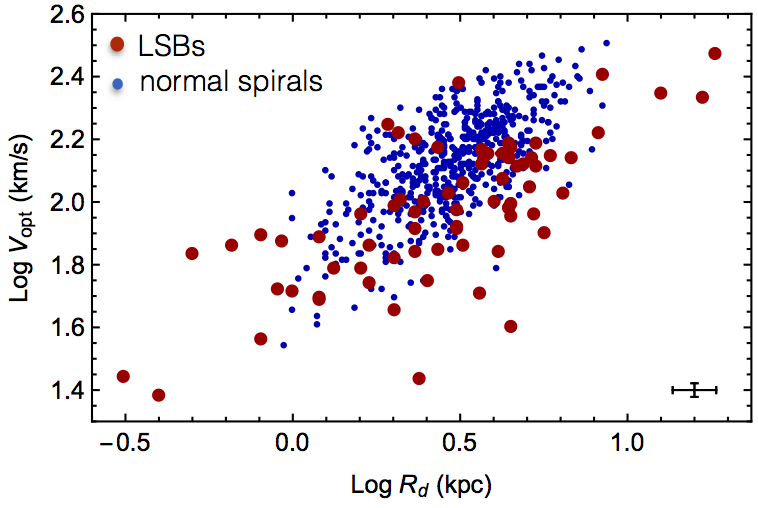}
\caption{Optical velocity $V_{\rm opt}$ vs. disc scale lengths in LSB galaxies (red) and in normal spirals (blue) \citep{Persic_1996}.
 The typical fractional observational uncertainties are 5\% in $V_{\rm opt}$ and 15\% in $R_{d}$, (see the right-down corner). Image reproduced {from} \citep{DiPaolo_2019}.}
\label{Log_Vopt_vs_Rd}
\end{figure}

LSBs galaxies are (obviously) characterised by low surface density exponential stellar discs \citep{deBlok_1996, Burkholder_2001, ONeil_2004}. Inside their 25 B magnitude isophotal radii, the stellar disk surface densities $<\Sigma_\star>$ are in the range (10 to 20) $\, M_{\odot}/pc^2$ (see Table 2 in \citep{Lei_2019}), values about 5--10 times smaller than those in HSB spirals of similar stellar disk masses. Noticeably, LSBs cover, in the range of their stellar disk masses $M_D$, the whole range found in spirals, from $\sim$10$^7 M_{\odot}$ to $\sim$10$^{11} M_{\odot}$ (see {Figures} \ref{MH_MD}, \ref{Previous_work_Md_Rd} and \ref{RC_RD_relation}). Similarly, their stellar disc scale lengths $R_D$ span from fractions of kpc to tenths of kpc (see, e.g., {Figure} \ref{Log_Vopt_vs_Rd}). Their magnitudes range as: $-22 \lesssim M_B \lesssim -10$ (see Table 2 in \citep{Du_2019}), $ -23 \lesssim M_R \lesssim -14$ (see Figure 2 in \citep{Minchin_2004}). 
A~detailed description of the photometric properties of LSBs can be found in \citep{Du_2019, O_Neil_2000, McGaugh_1994b, deBlock_1995} and a suitable comparison with the same properties in HSB galaxies is made in \citep{Du_2019}. 

\vspace{-6pt}

\vspace{-6pt}

LSBs are not rare objects; they likely comprise $\gtrsim$50\% of the general galaxy population (e.g., \citep{McGaugh_1995, Bothun_1997, Dalcanton_1997, Trachternach_2006, Greco_2018, Honey_2018}) with obvious cosmological implications (\citep{Bothun_1997,Minchin_2004}). However, the LSBs detection is challenging, due to their surface brightness that is much lower than that of their HSB counterparts and more difficult to detect against the sky \citep{Disney_1976, Williams_2016}; observational capability and selection effects inevitably lead to a bias which may jeopardise the understanding of their evolution.

The LSB disc galaxies show different morphologies (see Figures \ref{Many_morphologies_LSB} and \ref{UGC1378}, \citep{Honey_2016,Honey_2018}) from irregulars to spirals. They include both dwarfs and giant galaxies; the latter are sometimes composed of a HSB disc embedded in a larger LSB disc extended out to \mbox{$\simeq$100 kpc,} as in Malin 1 \cite{Bothun_1987, Impey_1989, Boissier_2016} (see Figure \ref{UGC1378}). Only a small fraction of LSBs shows a stellar \mbox{bar (\citep{Honey_2016},} see Figure \ref{UGC1378}). The largest LSBs usually have a central bulge (\citep{Das_2013}).
LSB galaxies rarely are catalogued as red objects (\citep{O_Neil_2000}) and in most of the cases are bluer than normal (HSB) spirals, with their ({B}--V) colour index lying in the range {[}0.49; 0.52{]}
, well outside the average value of (B--V) $\simeq 0.75$ found in HSB spirals \citep{McGaugh_1994b,Wyder_2009,deBlok_1996, Schombert_2014, Pahwa_2018, Du_2019}.
The LSBs photometry shows the following peculiarity: a lack of correlation between the central surface brightness and colour with other galaxies properties, as the disc mass, the luminosity, and the disc scale length (\citep{McGaugh_1994b}, Figure 6 in \citep{Bothun_1997}, Figures 8--11 in \citep{Pahwa_2018}). 
 
\begin{figure}[H]
\includegraphics[width=0.40\textwidth,angle=0,clip=true]{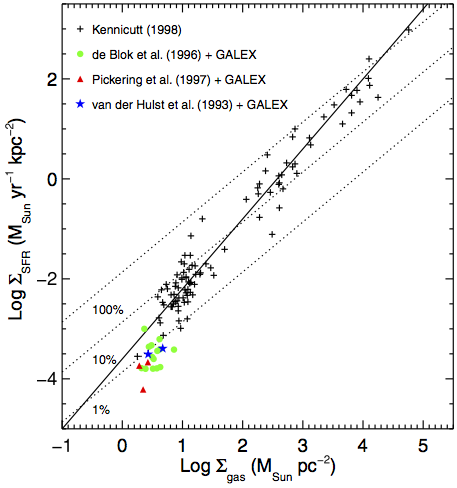}
\caption{Star formation rate surface density as a function of the HI surface density. LSBs from~\citep{Wyder_2009}. Gas surface densities from HI data: \citep{deBlok_1996} (green circles), \citep{Pickering_1997} (red triangles), and \citep{vanderHulst_1993} (blue stars). HSBs data (pluses) from \citep{Kennicutt_1998} with their power-law fit (solid line). The dotted lines show the relationship at a constant star formation efficiency of 100\%, 10\%, 1\%, in a timescale of star formation of $10^8$ yr. Image reproduced {from} \citep{Wyder_2009}.
 } 
\label{SFR_LSB_HSB}
\end{figure}

 HI radio observations indicate that LSB galaxies have very extended gaseous discs (out to several $R_{opt}$) with masses $M_{HI}$ ranging between $10^8$ and $10^{10} \, M_{\odot}$ ( \citep{O_Neil_2000, Pahwa_2018, Lei_2019}), i.e., of the order of their stellar disc masses $M_D$ (see {Figure} \ref{MH_MD} and Table 2 in \citep{Lei_2019}). In spirals the ratio $M_{HI}/M_D$, instead, varies from $10$ to $0.1$ along their magnitudes range. LSBs, then, show large values of $M_{HI}/L_B$ ratios (\citep{vanderHulst_1993, O_Neil_2000, Du_2019}) up to several times higher than those in normal spirals and range from $\simeq$0.1 to $\simeq$50 \citep{Burkholder_2001, ONeil_2004, Du_2019,O_Neil_2000}. One reason for such high values is the low density of their gaseous disks that prevents an efficient star formation~(\citep{Das_2009, Galaz_2011}) capable to turn the primordial HI disc in a stellar one as it occurs in HSB spirals. 
In fact, in LSB galaxies we find: $\Sigma_{HI} \sim 5 \,M_{\odot} pc^{-2}$ (see {Figures} \ref{fig6}\linebreak and \ref{SFR_LSB_HSB}, \citep{deBlok_1996, Lei_2019}) a value that is about half or less that in HSB galaxies of similar stellar mass (\citep{vanderHulst_1993}) and, therefore, according to the Kennicutt criteria \citep{Kennicutt_1989, Kennicutt_1998}, a value which is below the star formation threshold (\citep{vanderHulst_1993,Schmidt_1959, Kennicutt_1998, Boissier_2016}) implying that the gas is not ready to collapse and form stars \citep{vanderHulst_1993, Martin_2001, Blitz_2004, Robertson_2008, Wyder_2009}. In fact, the star formation rate (SFR) in LSBs is very low, usually $\lesssim$0.1 $\, M_{\odot} yr^{-1}$, i.e., at least one order of magnitude lower than in HSB spirals (\citep{deBlok_1996, van_den_Hoek_2000}, see also Table 3 in \citep{Lei_2018} and Table 2 in \citep{Lei_2019}).
In detail, typical values of the star formation surface densities are 
$$\Sigma_{SFR} \lesssim 10^{-3} M_{\odot} yr^{-1} kpc^{-2}$$ 
as shown in Figure \ref{SFR_LSB_HSB} and in Table 3 of \citep{Lei_2018}. The low star formation in LSBs also yields a low star formation efficiency (only a few percent that in HSBs) \citep{Lei_2018}. It is worth noticing that the LSBs have much lower SFR and $\Sigma_{SFR}$ than star-forming galaxies, despite both of them have similar HI surface densities (see Figure 10 in \citep{Lei_2018}). LSBs are characterised by a low metallicity (<1/3 of the solar value, see Figure 8 in \citep{McGaugh_1994b} and see also \mbox{references~\citep{Liang_2010, Bresolin_2015, Honey_2016}).} This causes an inefficient cooling with a consequent lack of large amounts of molecular gas and with a low dust content \citep{Matthews_2001, O_Neil_2003, Hinz_2007, Wyder_2009} that are important factors in determining the slow evolution of LSB galaxies. 
 The typically very blue colours of LSBs suggest that young stars are the dominant population, while the old stars do not make a substantial contribution to the light of the galaxy (e.g., \citep{Wyder_2009, Schombert_2014}). These properties, together with the observed low $H_\alpha$ emission (e.g., \citep{Schombert_2013}) and the high gas fractions, indicate a history of (very low) nearly constant with time star formation, compared to the exponential declining star formation of HSB spirals and irregulars (e.g., \citep{Vorobyov_2009, Schombert_2014}). 
Furthermore, the LSBs very low content of metals and dust, that are normally produced during the star formation process, also suggests that they formed relatively few stars over a Hubble time (see, e.g., \citep{Wyder_2009, Vorobyov_2009}). 

The LSB stellar population appears to be uniformly distributed in the stellar disc, since there is no significant colour gradient in the colour images \citep{Honey_2016}. Likely, the star formation is characterised by sporadic small-amplitude events (e.g., \citep{Schombert_2014}). 
Overall, LSBs are not the faded remnants of HSBs that have ceased to form stars as also suggested by the absence of any correlation between $\mu_0$ or colours and other galaxies properties (see, e.g.,~\citep{Bothun_1997, Pahwa_2018}). Rather, LSBs are slowly evolving galaxies separated from the normal spirals galaxies \mbox{(e.g., \citep{Vorobyov_2009, Schombert_2014})} and unique laboratories of astrophysics and cosmology.

 Let us anticipate that these astrophysical properties have a peculiar importance, in fact, although LSBs have very low ($\sim$0) SFRs and SFR surface densities over the whole Hubble time, they, remarkably, exhibit large core radii $r_0$ in the DM halo density, even larger than those of normal spirals that have undergone to a much higher star formation (see {Figure} \ref{Rho0_Rc}, upper panel). This evidence is in contrast with the idea that supernovae explosions (which are almost missing in LSB) are efficient dark halo core-forming processes in disk galaxies.

\begin{figure}[H]
\includegraphics[width=0.55\textwidth]{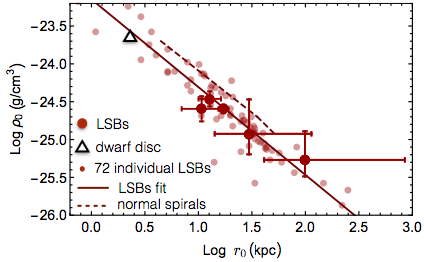}
\caption{\textit{Cont}.}
\label{Rho0_Rc}
\end{figure}

\begin{figure}[H]\ContinuedFloat
\includegraphics[width=0.55\textwidth]{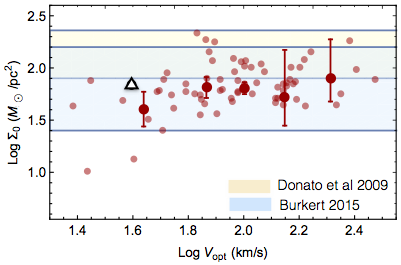}
\caption{(\textbf{up}): central DM halo density vs. core radius. Legend explains the details. (\textbf{bottom}): central surface density $\Sigma _0 = \rho _0 r_0$ vs. optical velocity $V_{\rm opt}$ (red points). Additionally shown the relation in~\citep{Donato_2009} (yellow shadowed area) and in~\citep{Burkert_2015} (light blue shadowed area). Dwarf discs data from~\citep{Karukes_2017}. Image reproduced {from} \citep{DiPaolo_2019}.}
\label{Rho0_Rc}
\end{figure}

\section{LSBs Mass Modelling. The URC Method}
\label{Results_from_LSBs}

 The structural parameters of the LSBs mass models are obtained by best-fitting their kinematics, more specifically, in \citep{DiPaolo_2019} they have been derived by means of the URC method applied to 72 RCs with 1614 independent $(r, V(R))$ measurements \citep{DiPaolo_2019}.\endnote{Online data link in \citep{DiPaolo_2019}.} The 72 RCs were selected from recent literature with the following criteria: (i) they extend out to \mbox{>0.8 $\,R_{\rm opt}$}, are symmetric, smooth and with an average internal uncertainty of <20$\%$; (ii) the galaxy disc scale length $R_D$ and the inclination function 1/sin $\,i$ are known within a $30 \%$ uncertainty. 
Although, in literature, there are investigations in which the above criteria are not considered, we stress here that they are, instead, necessary to correctly employ the disk kinematics to investigate the dark matter phenomenon. In the \citep{DiPaolo_2019} sample, the optical velocities $V_{\rm opt}$ span from $\sim$24 km/s to $\sim$300 km/s, covering the range of values of the full population of disk systems. Noticeably, the value of $V_{\rm opt}$ is not a LSBs discriminant (as for dwarf disks) contrary to the $V_{\rm opt}$ vs. $R_D$ relationship (see Figure \ref{Log_Vopt_vs_Rd}). In LSBs the latter is shallower as compared with that of normal spirals and with much larger internal scatter.

\begin{figure}[H]
\includegraphics[width=0.5\textwidth]{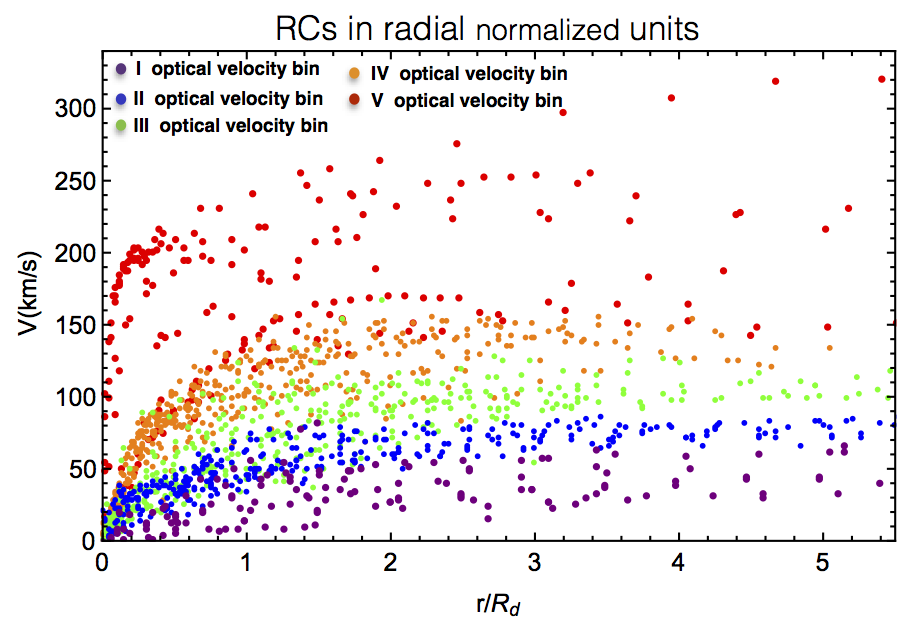}
\caption{The \citep{DiPaolo_2019} LSBs rotation curves (expressed in normalised radial coordinates) arranged in the I--V $V_{\rm opt}$ velocity bins and drawn in purple, blue, green, orange, red colour, respectively. Image reproduced {from} \citep{DiPaolo_2019}.}
\label{Vrnorm_Color}
\end{figure}

Following the URC method, ref. \citep{DiPaolo_2019} applied to each individual RC (a) the usual normalisations of the radial coordinate and the velocity amplitude and (b) the $V_{\rm opt}$ binning procedure. (see Sections 3 and 4 of \citep{Persic_1996}). The 72 RCs were, therefore, arranged in 5 $V_{\rm opt}$ bins according to their increasing values of such tag quantity, see Figure \ref{Vrnorm_Color}; in detail, according to the bin increasing Roman Number, we have: $ \langle V_{opt} \rangle = 43, \,73, \, 101, \, 141, \, 206 $ (in~km/s). Remarkably, after the double normalisation (DN), in each bin all the afferent RCs are very~similar.

\begin{figure}[H]
\smallskip 
\includegraphics[width=0.72\textwidth,angle=0,clip=true]{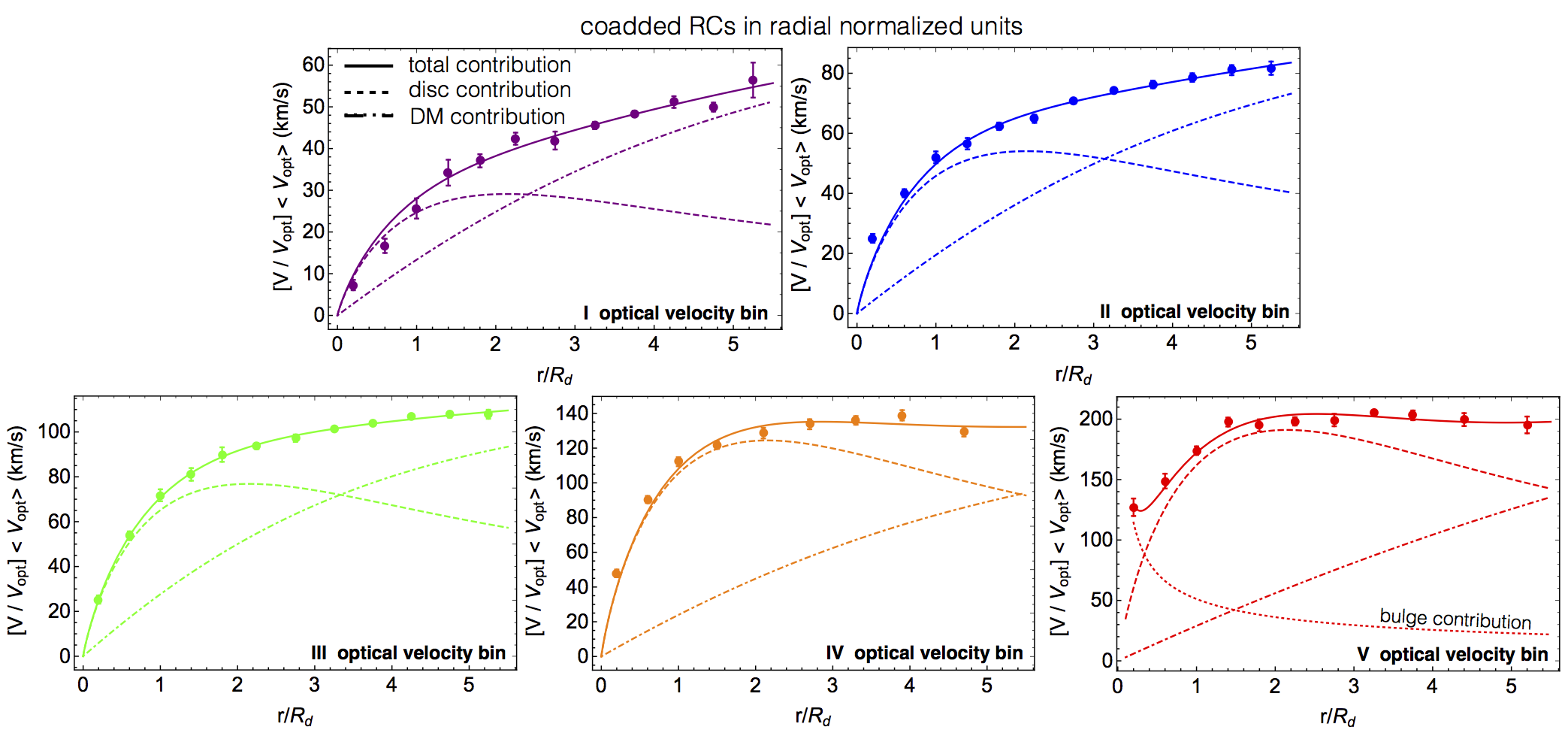}
\caption{The LSB-URC, i.e., the best-fit models (solid lines) of the coadded LSB-RCs (points with errorbars). The dashed, dot-dashed, dotted lines indicate the separate stellar disc, DM halo, stellar bulge contributions. Image reproduced {from} \citep{DiPaolo_2019}.}
\label{Velocity_grouped}
\end{figure}

\begin{figure}[H]
\includegraphics[width=0.5\textwidth]{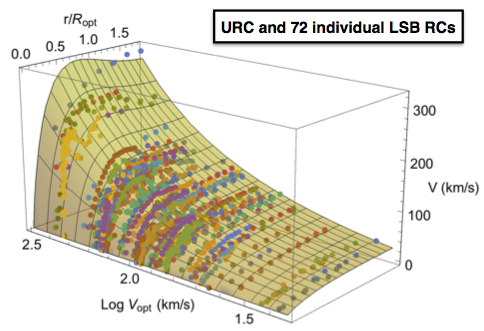}
\caption{URC-LSB (Equations (21) and (22) in \citep{DiPaolo_2019}), with compactness $Log \,C_\star=0$ and the individual 72 LSBs rotation curves. Image reproduced {from} \citep{DiPaolo_2019}.}
\label{URCpoints}
\end{figure}

 It is worth emphasising, also for LSBs, the advantages of the above procedure: in the 5 coadded RCs the peculiarities present in the individual RCs are smoothed out: the r.m.s. of the {\it coadded} RCs have been reduced down to (5--15\%), about half that of the individual RCs. 
 Then, by multiplying the 5 coadded DN RCs by the corresponding $\langle V_{opt} \rangle$ and $\langle R_{opt} \rangle$ values, one obtains the 5 coadded RCs in physical units (Figure \ref{Velocity_grouped}).
 They represent the whole LSBs kinematics ({Figure} \ref{URCpoints}) and show an universal trend with the quantity $V_{opt}$ that is analogous to that found in the URC of normal spirals (see Figure \ref{RC_Salucci})
\citep{DiPaolo_2019}. Furthermore, also for this family of disk systems the idea that {\it just one RC} can describe the dark matter phenomenon results plainly wrong and the claim according to which the RC curves are flat reveals itself just a fantasy.


\vspace{-12pt}

\vspace{-6pt}

These coadded RCs data are modeled, as in normal spirals \citep{Salucci_2007}, with an analytic function $V_{\rm URC}(R)$, chosen to be the sum in quadrature of the contributions from the Freeman stellar disc $V_d(R)$ (see Equation (\ref{V_stellar})) and from the DM halo $V_h(R)$. Well justified by the outcome of previous works on the mass distribution in galaxies, the latter contribution is assumed to originate from the {\it Burkert} halo profile (\citep{Burkert_1995, Salucci_2000,Salucci_20000}, see \mbox{Equations (\ref{V_dark_matter1}) and (\ref{DM_Mass})).} With this choice, by working out the best fit for the 5 $V_{\rm coadd}(R; V_{\rm opt})$, we determine the galaxy structural parameters. For the fifth $V_{opt}$ bin, we include also a stellar central bulge $V_{bu}$ \citep{Morelli_2012, Das_2013} (see Equation (\ref{Bulge_velocity})).


In first approximation, the inclusion in the velocity model of a HI gaseous disc component can be neglected \citep{DiPaolo_2019}. For the budget of the total baryonic matter in the LSBs the gaseous HI component is not negligible (see Figure \ref{UGC1378}), however, as this component is distributed at very large radii, the contribution $\simeq G M_{HI}(R)/R$ to the circular velocity, in the region where we have kinematics, is modest. Instead, outside the latter, the HI contribution overtakes the stellar one. The presently poorly known $M_D$ vs. $M_{HI}$ relationship is one of the most prominent goal of future LSBs investigations.

 The resulting baryonic fraction of the circular velocity is:
\begin{eqnarray}
\label{fb_definition}
f_b(R)= V_b^2 (R)/V^2 (R),
\end{eqnarray} 

Once we adopt the Burkert profile for the DM density, the total dark + baryonic contribution defines the candidate URC- LSB (see Equation (\ref{Bulge_velocity})).\endnote{In Equation (\ref{vur}), for simplicity, we have neglected the minor HI component}

\begin{eqnarray}
V^2_{URC}(R)=V_b^2(r;M_D,M_{bu}) +V_B^2(r;r_0,\rho_0)
\label{vur}
\end{eqnarray} 

As result the RHS of Equation (\ref{vur}) fits very successfully the 5 $V_{coadd}(r;V_{\rm opt})$, see \mbox{Figure \ref{Velocity_grouped}} and then establishes the URC-LSB. Furthermore, 
 the derived structural parameters $M_D$, $\rho_0$ and $r_0$ emerge as strong functions of $V_{\rm  opt}$ and $R_{\rm opt}$ (see Equations (7)\linebreak and (14) of \cite{DiPaolo_2019}).

\begin{figure}[H]
\smallskip
\includegraphics[width=0.59\textwidth,angle=0,clip=true]{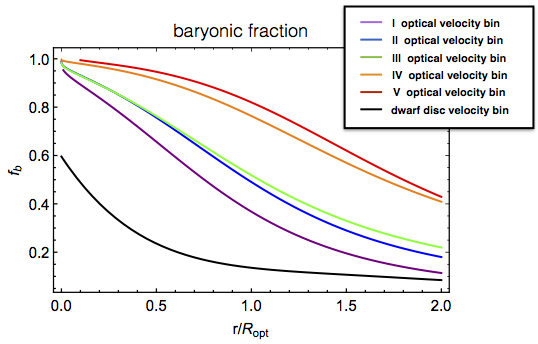}
\caption{Baryonic fraction $(V_b/V)^2$ as a function of $r/R_{\rm opt}$ derived from the URC-dd (black line, with $\langle V_{\rm opt} \rangle$ = 40 km/s) \citep{Karukes_2017} and from the LSB-URC (\citep{DiPaolo_2019}) evaluated at the $<V_{\rm opt}>$ values. Image reproduced {from} \citep{DiPaolo_2019_ggbx}.}
\label{fb}
\end{figure}

The baryonic mass fraction, as function of the {\it normalised radii} $R/R_{\rm opt}$ and the tag velocity $V_{\rm opt}$ is shown in Figure \ref{fb}.
From this latter we realise that, in the inner regions of the LSB galaxies, the stellar component is dominant, on the contrary, in the external regions, the DM component dominates. Moreover, the transition radius between these two regions increases directly with $R_{\rm opt}$ and $V_{\rm opt}$. A qualitatively similar behaviour is also observed in normal spiral galaxies (\citep{Persic_1996, Lapi_2018}).

Finally, the DM halos in LSBs turn out to be cored. First, let us stress again that the URC-LSB with velocity model including a Burkert {\bf cored} halo + the standard luminous components reproduces extremely well the coadded-LSB RCs. Secondly, ref. \citep{niloufar_2020} directly investigated the velocity model including a NFW {\bf cuspy} halo + standard luminous components by finding that this model is totally unable to reproduce the coadded data, as it occurs in normal spirals and dwarf disks \citep{Salucci_2007,Karukes_2017}.



 \section{Mass Modelling of Individual LSB Rotation Curves}
\label{Previous_results}

 For a complete investigation of the LSB structural relationships it is worthwhile to consider, also, their mass structures obtained in a number of works by modelling {\it individual high quality RCs}, (see \citep{DiPaolo_2019} and references therein). In the above, the mass model assumptions for the baryonic components were very similar to those described in the previous section and employed to reproduce the coadded RCs. For the DM halo, both the NFW profile and a cored halo profile, mostly the pseudo-isothermal one (see Equation \eqref{PISO}),\endnote{That, inside the inner galactic regions is in reasonable agreement with the Burkert profile (Equation (\ref{DM_density})) for $r_{0, B} \simeq 2 \ r_{0,pseudo-iso}$} were assumed.

In more than $90\%$ of the cases, the mass models with the cored DM profiles fit the circular velocities very well and in 50\% of the cases much better than the model with the NFW halo profile. ({see} \citep{deBlok_2002,DiPaolo_2020_PhD,Marchesini_2002, Swaters_2003, KuziodeNaray_2006, Kuzio_de_Naray_2008}). Furthermore, in the cases in which this model well reproduces a RC, the values of the best-fit parameters $c, M_{vir}, M_D$ result non-physical or in strong disagreement with (a) the photometric determination of the stellar disk mass of (b) the predictions of the $\Lambda$CDM cosmological simulations (\citep{deBlok_2002}, Figure 15 of \citep{Swaters_2003} and Figure 21 of \citep{Pickering_1997}). In the end, it is difficult to find LSB RCs whose (NFW halo + baryonic components) velocity model performs globally better than the corresponding that, instead, includes a cored DM halo.

\begin{figure}[H]
\smallskip
\includegraphics[width=0.69\textwidth,angle=0,clip=true]{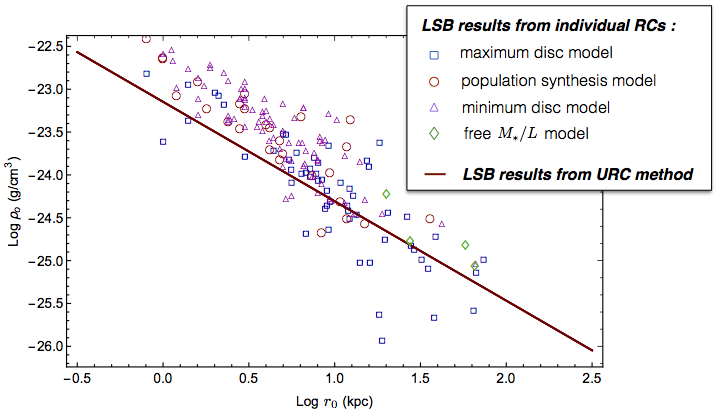}
\caption{Same legend of Figure \ref{Previous_work_Md_Rd}: the central density of the DM halo vs. its core radius. Image reproduced {from} \citep{DiPaolo_2020_PhD}.}
\label{Previous_work_rho0_Rc_Rd}
\end{figure}

In Figures \ref{Previous_work_Md_Rd} and \ref{Previous_work_rho0_Rc_Rd}, the values of the luminous and dark matter structural parameters, obtained from best fitting the individual RCs, are plotted alongside with those obtained by modelling the coadded LSB-RCs \citep{DiPaolo_2019} (see Section \ref{Scaling_laws}). 
Noticeably, the values of $\rho_0$, $r_0$ and $M_D$ determined, in the same object, by means of different data and different approaches, result in good agreement among themselves (\citep{DiPaolo_2019}, see Figures \ref{Previous_work_Md_Rd} and \ref{Previous_work_rho0_Rc_Rd}). The URC-LSB provide us with the systematics of the DM-LM coupling while the relationships from the individual RCs analysis provide us with an estimate of the internal scatter of such systematics.

 
\section{LSBs Structure Scaling Laws}

\label{Scaling_laws}
The URC-LSB mass models of the five coadded LSB RCs and of a reasonable number of high quality individual LSB RCs provide us with the structural parameters of the LSB galaxies. These, in turn, allow us to build up a series of relationships that characterise this family of disk objects \citep{DiPaolo_2019} and to compare them with those found in normal spirals \citep{Lapi_2018} and dwarf discs \citep{Karukes_2017}.

First, a particularly relevant relationship between the DM core radii $r_0$ and the stellar disc scale lengths $R_D$ emerges (see Figure \ref{RC_RD_relation}): 
\begin{eqnarray}
Log \, r_0 = (0.6 \pm 0.15) + (1.4\pm 0.25) \, Log \, R_D
\label{r0rd}
\end{eqnarray}with the values of these two quantities showing very large ranges: 12 kpc $\leq r_0 \leq$ 100 kpc, 1.2 kpc $\leq R_D \leq$ 8 kpc.

This relationship is consistent, caveat an offset of $ + 0.15 \pm 0.5$ dex, with the similar relationships found in spirals and dwarf disks. The physical nature of these two quantities tightly related by Equation (\ref{r0rd}) is intrinsically different: one defines the region in which the DM density is about constant with radius, the other establishes the exponential pace at which the disk surface luminosity declines with radius. They are also obtained in independent ways: the former is derived from the mass modelling of the galaxy kinematics, the latter, instead, is directly measured from galaxy photometry. Very remarkably, this relationship is also present in spirals and dwarf disks \citep{Persic_1996, Karukes_2017, Lapi_2018} and, therefore, highlights an amazing entanglement between the luminous and the dark matter in galaxies of different luminosity and morphologies. To propose that the relationship in Equation (\ref{RC_RD_relation}) be originated, rather than by the (interactive) nature of the dark matter particles, by some {\it astrophysical} process occurred in galaxies of very different luminosity and evolutionary history, seems an unsound and extremely fine-tuned idea.

\textls[-15]{The relation between the mass of the stellar disc and the optical velocity in} \mbox{Equation (\ref{RC_RD_relation})} is well known in disk systems as the bone of the Tully--Fisher relationship. In LSBs we confirm this by finding: 
\begin{equation}
Log \, M_D = (3.1 \pm 0.25) + (3.47 \pm 0.12) \, Log \, V_{\rm opt}
\end{equation}
with a rms of 0.24 dex \citep{DiPaolo_2019}. 
 Moreover, despite the very low SFRs, these galaxies show, for the same $V_{\rm opt}$ of spirals and dwarf disks, a (log) disk mass larger by 0.2 dex and 0.7 dex, respectively \citep{ Lapi_2018}. We must notice, however, that such stellar mass is distributed over an area about >4 times larger than that of the HSB spirals, this indicates the quantity $M_D/R_{\rm opt}^2$ as the discriminator between HSB and LSB galaxies.

%
%
%
%

Figure \ref{Rho0_Rc} (left panel) shows the relation between the DM halo central density and the core radius, which indicates that the highest densities are in the smallest galaxies as also found in normal spirals \citep{Salucci_2007}:
\begin{eqnarray}
\label{rho0_Rc_fit}
Log \, \rho _0 = -(23.15 \pm 0.07) - (1.16 \pm 0.05) \, Log\, r_0.
\end{eqnarray} 
(densities in g/cm$^3$, radii in kpc) with a rms of 0.2 dex. The LSB best fit line lies systematically 0.2 dex below the HSB one, probably this could be linked to a lower primordial DM density in the latter galaxies.
Moreover, the central surface density ($\Sigma _0$ expressed in units of $M_{\odot}/ pc^2$) follows the relationship (see right panel in Figure \ref{Rho0_Rc}): 
\begin{eqnarray} 
\label{Sigma_fit}
Log \, \Sigma _0 = Log \, (\rho _0 r_0) \simeq 1.9 \pm 0.2.
\end{eqnarray} 

 Thus, this relationship extends over 18 blue magnitudes over objects spanning from dwarf to giant galaxies (\citep{Spano_2008, Gentile_2009, Donato_2009, Plana_2010, Salucci_2012, McGaugh_2019, Chan_2019}) and very different morphology. It is difficult to not consider Equation (\ref{DM_density}) as a primary consequence of the dark particle properties, e.g., as it emerges in the case of a $\sim$2 keV neutrino (\citep{de_Vega_2014, de_Vega_2017}). 
 %


%
%

\begin{figure}[H]
\includegraphics[width=0.55\textwidth,angle=0,clip=true]{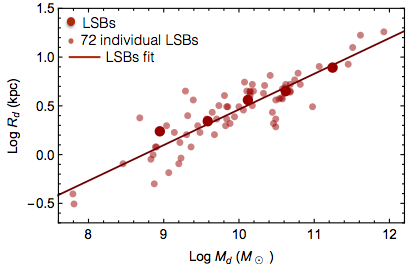}
\vskip 0.4cm
\includegraphics[width=0.55\textwidth,angle=0,clip=true]{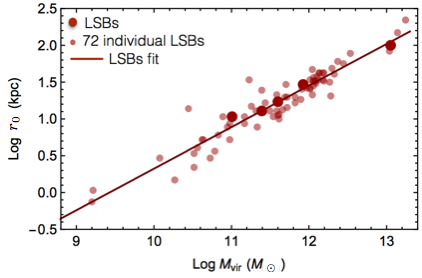}
\caption{(\textbf{up}) Stellar disc length scales vs. stellar disc masses. (\textbf{bottom}) DM halo core radii vs. their virial masses. Image reproduced {from} \citep{DiPaolo_2019}.}
\label{RD_MD}
\end{figure}

The two structural quantities of the stellar disc $M_D$ and $R_D$ correlate (Figure \ref{RD_MD}):
\begin{eqnarray}
\label{Stellar_Compactness_1}
Log \, R_D = (-3.19 \pm 0.23) + (0.36 \pm 0.02) \,Log \, M_D \quad 
\end{eqnarray}
with a rms of 0.24 dex, not a surprise according to the theory of the stellar disk formation in spirals. 
Furthermore, a correlation between the core radius $r_0$ (in kpc) and mass of DM halo $M_{vir}$ (in $M_\odot$) emerges (Figure \ref{RD_MD}): 
\begin{eqnarray}
\label{DM_Compactness_1}
Log \, r_0 = (-5.32 \pm 0.26) + (0.56 \pm 0.02) \, Log \, M_{vir} \quad.
\end{eqnarray} 
with a rms of 0.15 dex. This relation, theoretically presently unknown, is likely the effect of some property of the dark matter particles at microscopic level, such as, but not only, the presence of a quantum pressure in the innermost regions of the dark halos.

\begin{figure}[H]
\includegraphics[width=0.55\textwidth,angle=0,clip=true]{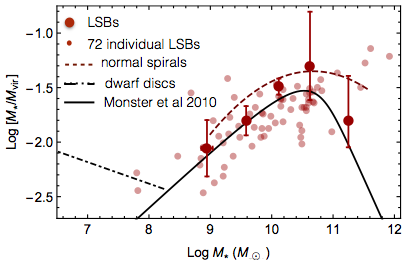}
\caption{Fraction of baryonic matter is LSBs vs. their mass in stars (points) compared with that of: normal spirals (dashed line) \citep{Lapi_2018}, various Hubble types (solid line) \citep{Monster_2010} and dwarf discs (dot-dashed line) \citep{Karukes_2017}. Image reproduced {from} \citep{DiPaolo_2019}.}
\label{M_star_over_Mvir}
\end{figure}

The total baryonic fraction is shown in Figure \ref{M_star_over_Mvir}: the lowest values are found in the smallest galaxies (and with the smallest stellar disc mass $M_D$). This ratio increases going towards larger galaxies and then reaches a plateau from which it decreases. This finding is in a certain agreement with the inverse ``U-shape'' of previous works relative to galaxies of different Hubble Types ~\citep{Lapi_2018,Monster_2010, Monster_2013}. This relationship reflects the 
effectiveness, over the whole Hubble time, with which the primordial HI associated to a galaxy, about 1/6 of its dark mass, has been transformed in stars. It is reasonable that it depends on many astrophysical effects and, on the side of the dark component, on its total mass.

\section{The Compactness}

In LSBs the structural mass parameters $\rho_0$, $r_0$, $M_{vir}$, $M_D$ show the same correlations among themselves found in normal spirals (see \citep{Yegorova_2007, Lapi_2018} and Figures \ref{Previous_work_Md_Rd} and \ref{Previous_work_rho0_Rc_Rd}) but all of them with a larger internal scatter ($\sim$0.35 dex). Motivated also by the fact that, at a fixed $M_D$, $Log \, R_D$ varies among the objects by almost $\simeq$1 dex (see Figures \ref{Log_Vopt_vs_Rd} and \ref{RD_MD}), ref. \citep{DiPaolo_2019} considered the presence in the URC-LSB, as in the URC-dwarf disks \citep{Karukes_2017}, of an additional structural parameter, namely the {\it compactness} $C_\star$. 
 Following \citep{Karukes_2017}, one defines $C_\star$ as the deviation between the value of $R_D$ ``expected'' from Equation (\ref{Stellar_Compactness_1}) given the galaxy mass $M_D$ and the value actually measured (see Figure \ref{RD_MD}).
 We have: 
\begin{eqnarray}
\label{Stellar_Compactness_2}
C_\star= 10^{( -3.19 + 0.36 \, Log \, M_D)} R_D^{-1},
\end{eqnarray} 
where let us remind that $R_D$ is measured from galaxy photometry while $M_D$ is derived from the URC mass modelling. By involving this additional quantity in the relationships expressed by Equations (\ref{r0rd})--(\ref{Stellar_Compactness_1}), their original scatter of $\simeq$0.35 dex gets reduced down to $\simeq$0.06 dex \citep{DiPaolo_2019}. Remarkably, two of these relationships involve only {\it structural quantities of the dark component}, however, as in dwarf disks, they become tighter when the (observational) quantity $C_\star$, belonging to the luminous world, is added in them as extra variable.
An other evidence of the entanglement of the dark and luminous matter in LSBs emerges when, in analogy with the quantity $C_\star$, one introduces $C_{DM}$, i.e., the {\it compactness of the DM halo} that describes the cases in which DM halos, with the same virial mass $M_{vir}$, exhibit different values for the core radius $r_0$. Then, by following Equation (\ref{Stellar_Compactness_1}), $C_{DM}$ can be written as:
\begin{eqnarray}
\label{DM_Compactness_2}
C_{DM}= 10^{( -5.32 + 0.56 \, Log \, M_{vir})}r_0^{-1}.
\end{eqnarray} 

Despite that collisionless DM halos and stellar disks cannot be pushed ones against the others, we find that the two compactnesses are positively correlated and in the same way in which that occurs in dwarf disks (at much smaller $V_{\rm opt}$) \citep{Karukes_2017}. We have, in fact \citep{DiPaolo_2019}:
\begin{eqnarray}
\label{DM_Compactness_3}
Log\; C_\star = (0.00 \pm 0.01) + (0.90 \pm 0.05) \,Log \, C_{DM}.
\end{eqnarray} 
with the small scatter of 0.15 dex. Again, we have a tight relationship between two quantities unrelated in the standard scenarios of DM: one deeply rooted in the dark world and the other in the luminous world. 

%

\begin{figure}[H]
\begin{center} 
\includegraphics[width=0.71\textwidth,angle=0,clip=true]{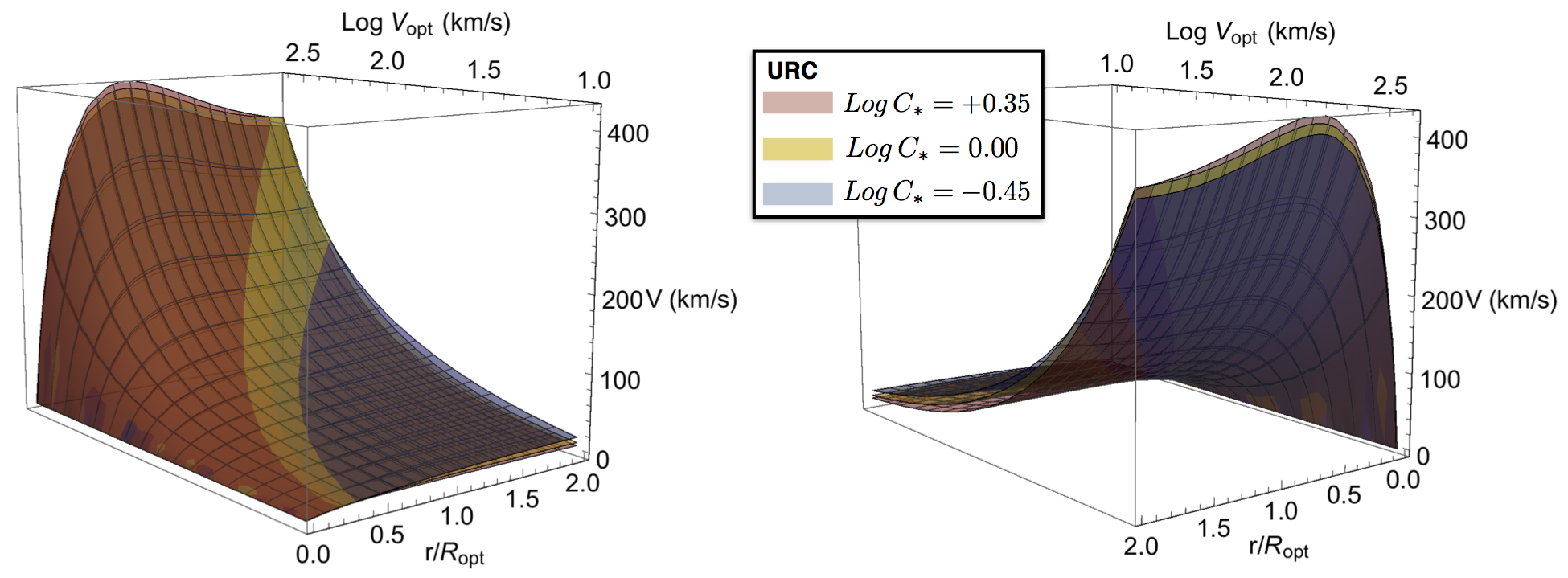}
\caption{URC-LSB in physical units for: low ($Log \,C_* = -0.45$), standard ($Log \,C_* = 0.00$) and high ($Log \,C_* = +0.35$) stellar compactness, respectively in {\it blue, yellow} and {\it red}.   Image reproduced from~\citep{DiPaolo_2019} (Figure 16).}
\label{URC_3D_Comp3_Vopt}
\end{center}
\end{figure}

With this new structural quantity $C_\star$ as a second parameter, one can build
 $$V_{URC-LSB}(R; R_{\rm opt},V_{\rm opt},C_\star)$$
(see \citep{DiPaolo_2019} for details and the 3D plot of this hyper-surface). Let us discuss the introduction in the URC-LSB $ V_{URC-LSB} (R;R_{\rm opt}, V_{\rm opt})$ of this tag quantity $C_\star$, i.e., of a second running observational parameter (see also \citep{Karukes_2017} for the case of the URC-{\bf dd}). The {\it one parameter} URC-LSB (see previous sections) fits reasonably well the 5 coadded LSB RCs, moreover, the structural quantities $\rho_0$, $r_0$, $M_D$ emerge all as functions of $V_{\rm opt}$ (and $R_{\rm opt}$). Then, the quantity: 
 $V_{URC-LSB}(R;R_{\rm opt},M_D(V_{\rm opt}),r_0(V_{\rm opt}),\rho_0(V_{\rm opt}))$ represents sufficiently well the individual LSB rotation curves.
 In details, the mean discrepancy between each of the 72 individual RCs and the corresponding ones predicted by means of the {\it one-parameter} URC-LSB via its $V_{\rm opt}, \, R_{\rm opt}$ values reads as:\endnote{In a sample, for the $j$th galaxy (with $V_{\rm opt}^j$ and $R^j_{\rm opt}$), the measured RC value at a radius $R^{ij}$ reads as: $V_{ind}^j(R^{ij})$} 
$$ <\Delta V/ \,V > \equiv \Big < \Big < \Big(\frac {V_{URC-LSB}(R^{ij},V_{\rm opt}^j,R_{\rm opt}^j)-V_{ind}^{j}(R^{ij})}{V_{ind}^j(R^{ij})} \Big)^2 \Big >^{1/2}_{{\rm over} \ ij } \Big>_{{\rm over }\, j }
 $$ 
Ref. \citep{DiPaolo_2019} {found}: $<\Delta V/ \,V > \simeq 0.19$, a relatively small value. Then, we introduce in the URC-LSB the additional dependence of the three parameters $\rho_0, r_0, M_D$ on {\it the observed quantity} $ C_\star$ and we get the {\it two parameters} URC-LSB: $ V_{URC-LSB} (R; V_{\rm opt}, R_{\rm opt}, C_{\star} )$ given by \mbox{Equations (20)--(22)} in \citep{DiPaolo_2019} (see {Figure} \ref{URC_3D_Comp3_Vopt}). Remarkably, with this addition, the discrepancy between the URC-LSB predictions and the 72 individual RCs data are reduced to\linebreak $<\Delta V/ \,V >\simeq 8 \%$, \textls[-15]{a value compatible with that of the URC-S and} \mbox{URC-{\bf dd}, \citep{Yegorova_2007,Persic_1996,Gammaldi_2018,Lopez_2018}} and in part due to observational errors in the RCs.

\vspace{-6pt}

In short, the disk compactness $C_\star$, arisen from the spread in the $V_{\rm opt}$--$R_D$ relationship (Figure \ref{Log_Vopt_vs_Rd}), is the main source of the scatter ($\sigma \simeq 0.35$) in the galaxy structure parameters scaling relationships (see Figures \ref{RC_RD_relation} and \ref{Rho0_Rc}) endowing so the URC with a second running~parameter.

\section{Angular Momentum}
\label{Angular_momentum}
 The derived mass structure of LSBs allows us to determine $j_\star$, the specific angular momentum (per unit mass) of their stellar component, that reads: (see \citep{Tonini, Rom_2012})\linebreak ($x \equiv R/R_D$),
 \begin{eqnarray}
 j_\star=f_R R_D V_{\rm opt} \hskip 0.9
cm f_R= \int^\infty_0 dx \, x^2 e^{-x} V_{URC}(x \, R_D,V_{\rm opt})/V_{\rm opt}\label{eq25}
 \end{eqnarray}
 
\begin{figure}[H]
\includegraphics[width=0.55\textwidth,angle=0,clip=true]{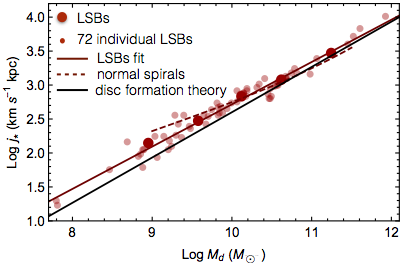}
\caption{LSBs stellar specific angular momentum-stellar mass (points) and its best fit (solid red line) compared with the spirals relationship (red dashed line) (\citep{Lapi_2018}) and with the relation $j_\star \propto M_D^{2/3}$ (black line) by \citep{Rom_2012}. Image reproduced {from} \citep{DiPaolo_2020_PhD} ({Figure 4.11}).}
\label{jstar}
\end{figure}

 In Figure \ref{jstar} we show the $j_\star$ vs. $M_D$ (km/s kpc vs. $M_\odot$) relationship: 
 \begin{eqnarray}
 Log \, j_\star = (-3.51 \pm 0.05) + (0.62 \pm 0.02) \,Log \,M_D
 \end{eqnarray}
 \vspace{-6mm}
 
This relation is in good agreement with that obtained in normal spirals \citep{Lapi_2018} and with the $j_\star \propto M_D^{2/3}$ relation for pure discs advocated by \citep{Rom_2012}. However, it shows an important property of the LSB galaxies.
The specific angular momentum of a DM halo $j_h$ is defined as: (see \citep{Mo_1998}) $$j_h= \sqrt{2} \lambda R_{vir} V_{vir}$$ where $V_{vir}^2= G \,M_{vir}/R_{vir}$ and $\lambda$ is the spin parameter of the DM halo, with an average value $\langle \lambda \rangle \approx 0.035$ which is nearly independent of mass and galaxy redshift (as indicated by numerical simulations: \citep{Barnes_1987, Bullock_2001, Maccio_2007, Zjupa_2017,Peirani}). We can compute for these objects the fraction $f_j$ of the primordial angular momentum per unit mass which is retained by the stellar disc after its building, 
$$
f_j= \frac{j_\star}{j_h}= 0.45-0.7
$$
and we find a value significantly smaller than $\simeq$0.8, the value found in spirals \citep{Lapi_2018}) characteristic of the case in which baryons and DM have conserved the primordial angular momentum per unit mass during the process of disk formation. This low value could be due to the fact that in LSBs the high angular momentum gas located in the outermost regions of the primordial HI {+} $H_2$ disks has been inhibited in transforming in stars by their very low surface densities (\citep{Fall_1983, Shi_2017}). In fact, if we consider the {\it baryonic}\endnote{The sum of the stellar and the HI} surface density and mass, one finds from Equation (\ref{eq25}): $
f_{j,b} \simeq 1.$


\section{ Accelerations in Low Surface Brightness Galaxies}
\label{LSBs_and_the_gravitational_acceleration}

The LSB galaxies, alongside with the dwarf disc galaxies, turn out to be crucial objects to investigate a possible universal relation between the radial gravitational acceleration $g(R)$ and its baryonic component $g_b(R)$ \citep{DiPaolo_2019_ggbx} as first claimed by McGaugh et al. \citep{McGaugh_2016}: (see below and Figure 3 of \citep{McGaugh_2016}). It is worth noticing that, despite that a $g$ vs. $g_b$ (see Section \ref{Galaxies_the_luminous_and_dark_matter_distribution}) relationship is searched, both accelerations depend on galactic radius, as can be seen in Equation (\ref{DM_gravity}); in each object one has: 
$$ 
g_b(R) = f_b(R )g(R), 
$$ 
where $f_b(R)$ is the baryonic fraction, function of $V_{\rm opt}$ and $R$ (see Figure \ref{fb}). McGaugh et al. \citep{McGaugh_2016} have stressed that $g(R)$ shows a very surprising feature: it correlates, at any radius and in any object, with the ``baryonic'' $g_b(R)$ and this in a way {\it very different} from the $g = g_b$ relationship expected in the no-DM Newtonian case. In detail, their ($g_b$, $g$) data are fitted~by:
\begin{eqnarray}
\label{g_McG+16}
g(R) = \frac{g_b(R)}{1-exp \left(-\sqrt{\frac{g_b(R)}{\tilde{g}}} \right)},
\end{eqnarray} 
with $\tilde{g}= 1.2 \times 10^{-10} \,$ m s$^{-2}$, see Figure 3 in \citep{McGaugh_2016}. At high accelerations, $g \gg \tilde{g}$, \mbox{Equation (\ref{g_McG+16})} converges to the Newtonian relation $g = g_b$ while, at lower accelerations, $g < \tilde{g}$, Equation (\ref{g_McG+16}) strongly deviates from the latter (\citep{McGaugh_2016, McGaugh_2018}). This relationship with a claimed internal scatter of only 0.13 dex seems to bend towards the Milgrom dynamics rather than to the standard Newtonian DM scenario. Let us stress that the $g$--$g_b$ relationship is almost all-observational: $g$ comes fully from observations while $g_b$ comes also from observations and from adopting a method to derive the disk mass from the latter, that is expected to induce a negligible bias in the relationship of Equation (\ref{g_McG+16}). Moreover, the observational errors in the quantities used to estimate $g$ and $g_b$ have small effect in~the~latter.

Concerning spiral galaxies, ref. \citep{Salucci_2018} confirmed and statistically extended the above results. Interestingly, he assumed the presence of DM halos as the origin of the ``anomalies'' in the accelerations and used 100 K accelerations measurements from about 1200 spirals. The disk masses (and then the values of $g_b(R)$) were obtained from the URC-S and from the radial Tully--Fisher relation (\citep{Yegorova_2007}) in ways alternative to the spectro-photometric method of \citep{McGaugh_2016, McGaugh_2018}. Thus, in the \citep{Salucci_2018} approach the presence of the DM halo is explicit while in~\citep{McGaugh_2016, McGaugh_2018} the approach is agnostic to such presence. The outcome, see Figure 13 in \citep{Salucci_2018}, is a relation among the two accelerations quite consistent with those in \citep{McGaugh_2016, McGaugh_2018}, albeit with a larger scatter of 0.25 dex, due also to a mild dependence of the \citep{Salucci_2018} $(g,g_b)$ pairs on the disk masses. The fair agreement (in spirals) between the previous studies indicates that both the assumption of a DM halo or any reasonable estimate of the stellar disk mass do not bias the two-accelerations relationship.

\begin{figure}[H]
\includegraphics[width=0.62\textwidth,angle=0,clip=true]{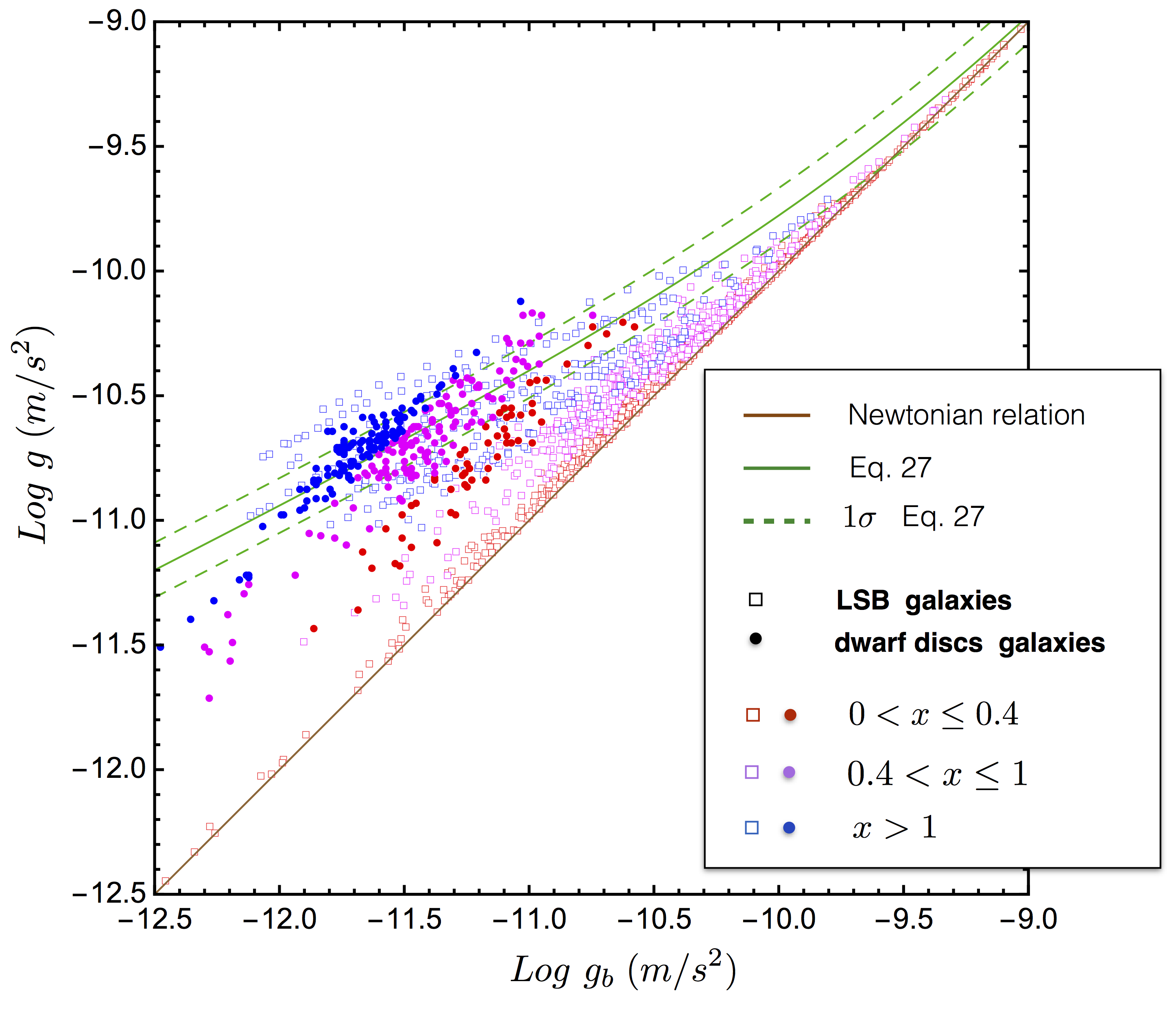}
\caption{Total acceleration $g$ vs. its baryonic component $g_b$. $x\equiv R/R_{\rm opt}$. Additionally shown: Equation (\ref{g_McG+16}) (green line) and its $1\, \sigma$ error bar (dashed green lines) and the Newtonian relationship: $Log \, g = Log\, g_b $ (brown line). Image reproduced {from} \citep{DiPaolo_2019_ggbx}.}
\label{ggb}
\end{figure}

Reference \citep{DiPaolo_2019_ggbx} realised that the $g(g_b)$ relation, holding in normal spirals breaks down in LSBs and dwarf disks, see Figure \ref{ggb} (\citep{DiPaolo_2019_ggbx}). The failure is due: (i) any proper relationship between $g$ and $g_b$ necessarily must involve also the (normalised) radius $x=R/R_{\rm opt}$ where the two accelerations are measured; (ii) in disk systems mass models tell us that the fraction of baryonic matter is a complex function of $x$ and $V_{\rm opt}$. The McGaugh et al. relationship, with only two quantities involved, cannot follow the complex distribution of luminous and dynamical mass in galaxies of different luminosity and mass. We remark that in spirals, since the DM dominance is less prominent and the range of variations of $V_{\rm opt}$ and $C_\star$ are smaller than in LSBs, (i) and (ii) affect much lesser the relationship and their effects are hidden within its big scatter.

\begin{figure}[H]
\smallskip
\includegraphics[width=0.62\textwidth,angle=0,clip=true]{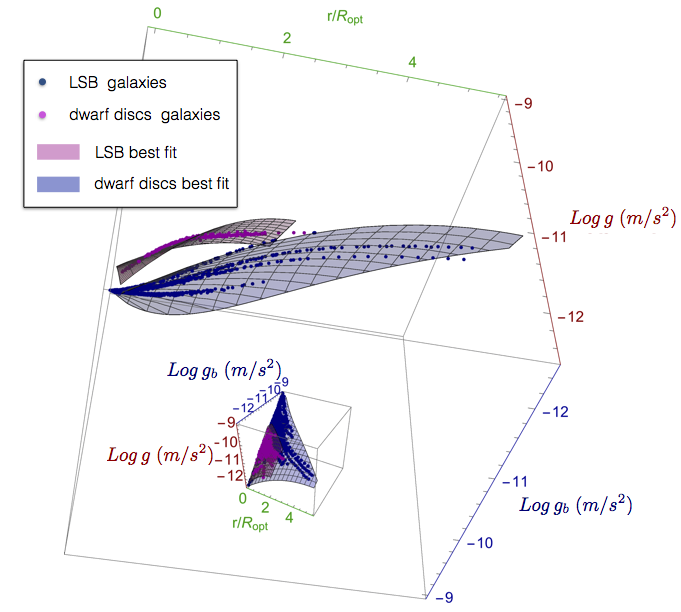}
\caption{Total acceleration $g$ vs. baryonic acceleration $g_b$ vs. normalised radii $r/R_{\rm opt}$. The magenta and blue points are the {\bf dd} and LSB data. The magenta and the blue surfaces are the {\bf dd} and LSB best fitting models. Image reproduced {from} \citep{DiPaolo_2019_ggbx}.}
\label{ggbx}
\end{figure}

The actual relationship that realises both the McGaugh et al. underlying idea and the entangled distribution of mass in galaxies has been devised in \citep{DiPaolo_2019_ggbx} and shown in \mbox{Figure \ref{ggbx}} (hereafter $gg_bx$ {relation}). The surfaces drawn are {\it empirical} analytical functions that best fit the data. The scatter of data around them is only 0.05 dex, i.e., 1/6 of that which the same data show around the McGaugh et al. relationship. This is extremely remarkable, implying a tight relation linking the total and the baryonic accelerations, the galactocentric normalised distance $x=R/R_{\rm opt}$ and the morphology of galaxies.

In the Newtonian Gravity paradigm, the amazing $gg_bx$ relationship indicates the presence of a complex entanglement between the dark and luminous mass components. Let us notice, however, that the simplest realisation of this new {acceleration radius} 
relationship, likely indicating the underlying physics, is still to be determined.

\vspace{-12pt}
%

\section{A Direct Interaction between Luminous and Dark Matter from the Structural Properties of the LSBs?}
\label{LSBs_toward_new_hints_for_DM}

The analysis of the matter distribution in galaxies leads us to realise the profound interconnection that is present in them between the luminous and the dark components. In LSBs galaxies all this is even more spectacular: the two (diverse) components, indeed, are linked by means of tight scaling relationships among their structural properties. First, the $r_0$--$R_D$ relationship (Figure \ref{RC_RD_relation}) emerges also in LSBs as in spirals and in dwarf disks. This is of extraordinary relevance: the DM core radius $r_0$, whose cosmological creation is still unexplained, and the stellar disc scale length $R_D$ result entangled despite being intrinsically very different quantities and derived in totally independent ways. This seems very difficult to occur in a collisionless DM scenario.

Remarkably, LSBs have an important role in upgrading such difficulty in impossibility. In fact, while in spirals one could argue that some very fine-tuned baryonic feedback from supernovae explosions has transformed the originally cuspy halos into the actual cored ones and, as a  byproduct, has created the above relationship (see, e.g., \citep{Dutton_2019}), in LSBs, instead, we cannot entertain us with such a fantasy in that these objects have had a very low SFR over their entire life that has not allowed the formation of the required feedbacks. In addition, there are giant LSBs with masses $ M_{vir}\sim 2 \times 10^{13} M_\odot$ and virial radii of about 700 kpc, that have strong gravitational fields able to prevent the dynamical effects of the supernovae feedbacks on the DM halo density, but still show giant dark halo core radii as wide as about $60$ kpc. 

Central supermassive black holes have been claimed able to create cored DM density distributions, however, it is well known that in LSBs the SMBHs, if present, are very small: their RCs can give upper limits to these masses: by applying the method of \citep{Salucci_2000,Salucci_20000} for the LSBs of \citep{DiPaolo_2019} one finds:$$M_{SMBH-LSB}< 3 \times 10^6 \ \Big(\frac{V_{\rm opt}}{50 \ km/s}\Big)^2 \ M_\odot$$
that implies that these central condensations can gravitationally influence a region no wider than (100--300)$\ pc << r_0(V_{\rm opt})$. 

The above discussed entanglement between the dark and the luminous mass distribution emerges in the following relationships: 

\begin{itemize}
\item  $log \ r_0$ vs. $ log \ R_D$; and 
\item $log \ g$ vs. $log\ g_b$ vs. $log \ x$, holding for disk systems; and 
\item  $log\ C_{DM}$ vs. $log \ C_\star$, holding for LSBs and dwarf disks.
\end{itemize}

See \citep{Salucci_2019} for a more detailed review. Furthermore, the entanglement is also present in the relationships:

\begin{itemize}
\item $log \ \rho_0$ vs. $ log \ M_D$ (\citep{Salucci_2007}); and 
\item  $\Sigma_\star(r_0) =const \ $ (\citep{Gentile_2009}) holding for spirals;
\item $\rho_0 r_0= const $, (\cite{Salturini,Donato_2009}, and references therein) holding for the dark world of \linebreak all~galaxies.
\end{itemize}

Considering also the lack of detection of a collisionless DM particle via direct, indirect or collider methods, ref.  \citep{Salturini} proposed the existence of a {\it direct interaction} between the dark and the luminous matter components. This can be represented, for{\it a} halo of mass $M_{vir}$, by the following evolutive equation for the DM density:
\begin{eqnarray}
 d \rho_{DM}(r,t)/dt =- I(\rho_{DM}(r,t),\rho_{L}(r), <v_{DL}>)
\label{int}
\end{eqnarray}
where $\rho_{DM}$ is the DM density as function of radius and time,   $<v_{DL}>$ and $\rho_L$ are the average relative velocity between dark and luminous particles\endnote{More specifically: any SM particle.} and the density of the latter, both function of radius and constant with time. In this way, the evolution and the state of the DM density depends on its initial condition $\rho_{DM}(r,0)$, but also on the distribution of luminous matter $\rho_L(r)$. The existence in Equation (\ref{int})  of a dark-luminous  coupling  can trigger an entanglement among  the structural properties of the two different components. Of course, in order to investigate the whole family of disk systems, \mbox{Equation (\ref{int})} has to be solved  after inserting in it  the   $M_{vir}$ dependence  in all its terms.  

\begin{figure}[H]
\smallskip
\includegraphics[width=0.68\textwidth,angle=0,clip=true]{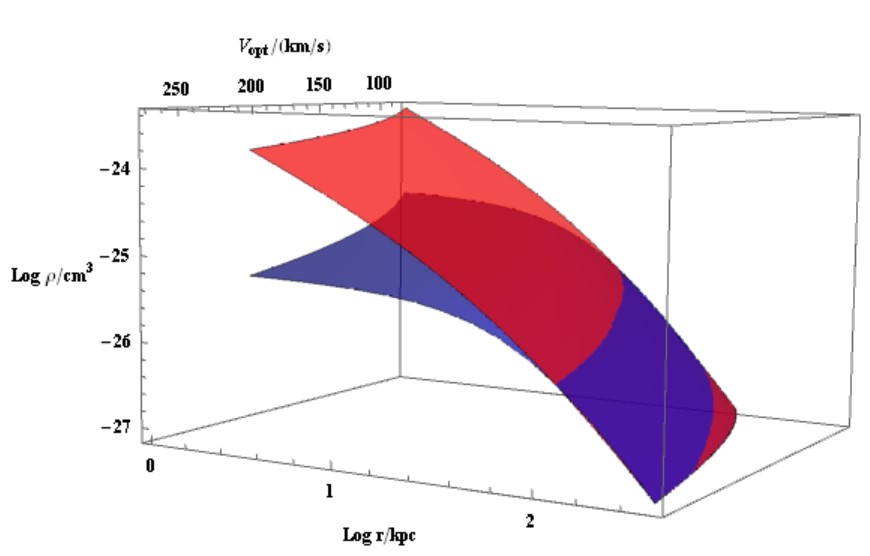}
\caption{Primordial (red) and present (blue) DM density profiles as a function of radius $r$ and of optical velocity $V_{\rm opt}$. Image reproduced {from} \citep{Salucci_2007}.}
\label{Cusp_core}
\end{figure}

In other words: in the inner luminous parts of galaxies, the DM particles, on a Hubble timescale, exchange energy with the standard matter particles in a way other than via gravity.  This exchange creates the DM cores and  the detected luminous-dark matter entanglement.  To work out the details of such DM--SM particle interaction is a main goal for the future investigations on the nature of the DM particle. In addition to the previous indirect support, there is also a direct imprint of a DM--SM particles interaction occurring on a timescale of $\sim$10$^{10}$ yr. Let us stress that, also in a collisional DM scenario, the dark halos are formed (at high redshifts) within a free fall time of $10^{7-8.5}$ yr, i.e., in a time much smaller than the collisional one and then in a collisionless way  that yields to the profile of Equation (\ref{NFW_profile}) \citep{Navarro_1996}. 
Such a feature is what we recover in the outermost regions of the $z=0$ dark matter halos. In fact, in all disk systems \citep{Salucci_2007,Karukes_2017, DiPaolo_2019} found that, for $r>r_0$, i.e., outside the region in which the collisional interactions have mostly taken place in the past 10 Gyr and smoothed out the primordial cuspy DM profile, the DM halo densities are well reproduced by the {\it  collisionless} profile $\rho_{NFW}(r;c(M_{vir}),M_{vir})$ (see Figure \ref{Cusp_core}), with: 
 $$c(M_{vir})\simeq  14 \ (M_{vir}/(10^{11}M_\odot))^{-0.13}$$ 

\vspace{-24pt}
 
Equation (\ref{NFW_profile}), then, reproduces the density distribution of the DM halo {\it before} \textls[-25]{that the DM-LM interactions took place, so that, one can} set: $\rho_{DM}(0,r,M_{vir})=\rho_{NFW}(r, c(M_{vir}),M_{vir})$ where we have  made explicit the dependence of the initial DM density on the concentration. 

The amount of dark matter that,  in a Hubble time, has been removed from the central region out to $r_0$ is: 
\begin{eqnarray} 
 \label{eq16}
\Delta M_{DM}(r_0)= 4 \pi \int^{r_0}_0 (\rho_{NFW}(r, M_{vir}) - \rho_{B}(r,M_{vir})) r^2 dr.
 \end{eqnarray}

 In all disk systems this amount ranges from 40\% to 90\% of the value of $M_{NFW}(r_0)$, the primordial mass inside $r_0$ and, very remarkably, it is just $\frac{1}{100}$ of the present and, therefore, of the primordial value of halo mass $M_{vir}$. In all cases, the core forming process has not changed the mass of the DM halo, although, has largely removed the dark particles from its innermost regions <$r_0(M_{vir})$. Next step is to estimate the number of interactions and the exchanged energy in the core forming process that has flattened the primordial cuspy dark halo density. Given $m_p$ the dark particle mass, the number of interactions per galaxy involved in the process is: 
$N_I(M_{vir})=\Delta M_{DM} (M_{vir})/m_p$. The 
number of interactions for galaxy atom of mass $m_H$ is 
\begin{eqnarray} 
N_{I/A}(M_{vir}) = \frac{\Delta M_{DM}(M_{vir})}{M_\star(M_{vir})} \,m_H/m_p. 
\end{eqnarray}

$W(M_{vir})$, the work done during the core-forming process, is obtained by: 
\begin{eqnarray}
\nonumber
 W(M_{vir})= 4 \pi \ \Big( \int_0^{r_0(M_{vir})}\rho_{NFW}(r;M_{vir}) M_{NFW}(r;M_{vir})\,r \, dr - 
 \\
 \int_0^{r_0(M_{vir})} \rho_{B}(r;M_{vir}) M_B(r;M_{vir})\, r\, dr \Big)
\end{eqnarray}

 Following \citep{Salturini}, we divide this energy by the number of interactions $N_I(M_{vir})$ taken place in each galaxy inside $r_0(M_{vir})$ during the Hubble time and we get the energy per interaction and per GeV mass of the dark {particle}: $E_{core}=(100-500) \ eV \ \frac {m_P}{GeV} $.
 
It is worthwhile to discuss more specifically about the scenario we have brought out for attention: we envisage at least two possibilities for the postulated interaction: (a) an increase in the particle self-annihilation in dense baryon dominated environments like, e.g., stars or (b) a scattering, in the same objects, between dark-luminous particles that captures the former or remove it from the inner region of the galaxy. Noticeably, at the center of the Sun, certainly not one of the densest stars, standard model particles have a temperature of about 2 keV sufficient to trigger an effective interaction.  

\section{Conclusions}

After reviewing the previous observational evidence one could argue whether the above ``entanglement'' could arise only from the dark matter particle properties in a standard astrophysical scenario or must imply a new ``interaction'' between dark particles and luminous ones. Both possibilities are totally excluded for the popular primordial black hole DM scenario, in this $\Lambda CDM $ scenario we cannot envisage a mechanism of core-forming. The first possibility is quite in jeopardy for the WIMP scenario for which Equation (\ref{int}) does not hold, and the baryonic feedback struggles to reproduce the whole observational data. In the ULA and in the standard SIDM scenarios, cored distribution emerge from their DM physics but with sizes in strong disagreement with observations: in the first case they {\it decrease} with halo mass \citep{Burkertula}, in the second they {\it do not depend on it}. Only a (fine tuned) velocity dependent SIDM scenario might reproduce Equation (\ref{Sigma_fit}) but also in this case we do not see how an entanglement between dark and luminous matter might~arise. 

A sort of dark-luminous matter coupling could occur in the 2 keV WDM particles scenario. Such a  particle has, at galactic scales, a quantum pressure whose equation of state depends also on the distribution of the baryonic matter, namely the stellar disk. Therefore, it is not totally surprising that the predicted galaxy structure, in the case of a self-gravitating \mbox{2 keV} fermionic particle, be in agreement with the Spiral and Dwarf Disks URCs \citep{ de_Vega_2013,de_Vega_2014, de_Vega_2016} which is a good starting point to cope with the above observational entanglement. On the other hand, it is well known that cosmological properties of the Lyman $\alpha$ clouds at intermediate redshifts and other cosmological observations give an upper bound of the mass of the WDM particle of >3 keV (e.g., \citep{Vegetti_2018,Enzi_2020}) relevantly higher than that required to form cores in galaxies. However, such mass is the thermal relic one (i.e., the mass of the WDM particle if it decouples from the expanding Universe in thermal equilibrium) which may be not the case for the sterile neutrinos. The relation between the physical particle mass of the \mbox{$\sim$keV} neutrino in galaxy halos and the corresponding thermal mass, which is cosmologically tested, has to be worked out explicitly for every specific WDM particle physics \mbox{models \citep{merle_2012,Merle_2013,devega2012}} many of which, however, have been already \mbox{discarded \citep{Alvey_2021}.} Finally, it is still under study the role that, in this scenario, the standard galactic astrophysics may take by triggering the observed DM-LM entanglement.

In the past 40 years, astrophysical observations have been overwhelming pointing to the existence, in virialized objects, of a large amount of matter, different from the standard luminous (baryonic) matter and dubbed as dark. However, many puzzles around such ``DM phenomenon'' are still unresolved. In this review we have presented the astrophysical properties  of the DM in galaxies related with the intrinsic properties of the dark particle by focusing on low surface brightness galaxies. These objects, very numerous in the Universe, are unique cosmological laboratories for investigating the dark matter phenomenon. In fact, their stellar and gaseous components are the most extended  in galaxies and provide us with an equally very extended kinematics. In addition, given their very low Star Formation rates and stellar surface densities, these galaxies do not develop a significant baryonic feedback, as it occurs in galaxies of different Hubble types. The investigation on the dark component is, therefore, much simplified. It is worth noticing that, in spite of the many differences that LSBs show with respect to the other disk systems, their stellar disk angular momentum per unit mass correlates with disk mass in a way very similar to that found in normal spirals (of high surface brightness). The tag of a LSB galaxy, therefore, is not its mass, size, or angular momentum but its $\frac {mass} {surface}$ ratio.

As all other disk systems, LSB galaxies have been investigated by means of the universal rotation curve method which allows one to derive, for each system (e.g., normal spirals), a Universal kinematics and then an Universal mass structure which depend on few galactic parameters as the optical radius $R_{\rm opt}$ and the optical velocity $V_{\rm opt}$. We have reviewed the study \citep{DiPaolo_2019} in which a sample of 72 individual RCs has provided us with 5 coadded RCs from which the URC-LSB has been determined. This latter includes the contributions to the circular velocities from the standard Freeman stellar disc and from a Burkert cored dark halo, exactly as in URC-S and URC-DD. The three free parameters of the URC-LSB: $\rho_0$, $r_0$, $M_D$ emerge all as a function of $V_{\rm opt}$. As result, $V_{URC-LSB}(R; R_{\rm opt},\rho_0(V_{\rm opt}), r_0(V_{\rm opt}), M_D(V_{\rm opt}))$ well describes the individual rotation curves of this family of galaxies: the resulting average percent error in predicting them is only $\Delta V/V \simeq 14 \% $ and it reduces by a factor two when in the URC-LSB we add the new parameter $C_\star$, (see \citep{DiPaolo_2019}) the compactness related to the spread of the $V_{\rm opt}$--$R_D$ relationship (Figure \ref{Log_Vopt_vs_Rd}).

The URC and the investigation of individual RCs of LSBs provide us with tight scaling laws among the luminous and the dark matter structural properties, as previously found for all the other disk systems \citep{Persic_1996, Karukes_2017,Lapi_2018}. Among these, one should highlight: (1) the relationship involving the stellar disc scale length $R_D$ and the DM core radius $r_0$ and (2)~Equation~(\ref{rho0_Rc_fit}) involving the DM halo central density $\rho_0$ and the DM halo core radius $r_0$.

Furthermore, a new parameter, the concentration $C_\star$ helps in fully describing the kinematics of LSBs, as well as that of DDs \citep{Karukes_2017}. The dependence of the galaxy scaling laws on this new quantity (that adds up to the quantities $R_{\rm opt}$ and $V_{\rm opt}$) gives rise to a new challenge for the $\Lambda$CDM N-Body + Hydro-dynamical baryonic feedback  effects scenario, in that we detect such entangled relation in   objects where the latter does not exist. 
 
\textls[-20]{It is worthwhile to point out, also, that these scaling laws found in LSBs, e.g., \mbox{Equation~(\ref{rho0_Rc_fit})}} seem very difficult to arise also within scenarios alternative to $\Lambda$CDM, such as ULA, SIDM and the popular $\Lambda CDM$ 30 $M_\odot$ primordial black oles scenario (see \citep{Capela_2013, Zumalac_2018, Niikura_2019}). Among other inconsistencies, in the latter scenarios, the dark `particle' seems totally unable to form the observed core radius vs central density relationship and the detected DM halo cores of a size of about 100 kpc. One exception could be the 2 keV WDM fermion scenario in which the DM quantum pressure depends also on the distribution of the luminous matter.
 
It has been useful to investigate in LSBs the relation between the gravitational acceleration $g$ and its baryonic component $g_b$ claimed by \citep{McGaugh_2016}. Considering also the outcome of a similar investigation in dwarf discs, ref. \citep{DiPaolo_2019_ggbx} realised that, in order to build a physical suitable relationship with the two accelerations, one has to involve also the normalised galactic radius $x\equiv r/R_{\rm opt}$ at which $g$ and $g_b$ are evaluated. This leads to a (new) relationship with a very smaller intrinsic scatter that highlights a strong entanglement between the dark and the luminous matter. 
Such observational evidences, plus the fact that (so far) the WIMP particle is undetected, lead one to strongly consider the existence of a direct LM-DM interaction (in addition to the gravitational one). This interaction is a key point for solving the mystery of the DM phenomenon.
 
Further studies are planned in order to clarify open issues inherent to LSBs and proceed into the investigation of the DM scenario: in particular, one needs:

\begin{enumerate} 

\item[(a)] To enlarge the LSBs rotation curves sample and increase their level of spatial resolution to have a better knowledge of the properties of these galaxies and of the various LM vs DM relationships. A larger statistic will also allow us a better approach of the URC method, by involving the compactness $C_\star$ from the beginning of the rotation curves~analysis;

\item[(b)] To study the giant LSBs, special objects which are often made of a HSB disc embedded in a large LSB disc. Dwarf and giant LSBs have different evolutionary histories \mbox{(e.g., \citep{Matthews_2001b})} and, moreover, we want to understand how the DM phenomenon realises itself over a range, for the halo mass, of 5 dex;

\item[(c)] To analyse the LSBs with strong peculiarities: very red objects (e.g., \citep{Burkholder_2001}); objects with near solar metal abundances \citep{Bell_2000}; giant objects with properties different from the average LSBs (e.g., \citep{Boissier_2016}), objects with bulge or a central AGN (e.g., \cite{Mishra_2018});

\item[(d)] \textls[-20]{To understand the reason (systems isolation or low values of the spin parameter~\cite{Dalcanton_1997b, Boissier_2003, DiCintio_2019}}) of the lower gas surface density in LSBs;

\item[(e)] To understand the systematic difference $\simeq$0.2 dex between most of the structural relationships found in LSBs and the corresponding ones in normal spirals (e.g., \linebreak Figures \ref{RC_RD_relation}, \ref{Rho0_Rc} and \ref{M_star_over_Mvir});

\item[(f)] To envisage observations in LSBs (as well as in other Hubble types) that could further reveal the presence of a LM--DM particle interaction;

\item[(g)]Tto obtain kinematical observations at high redshifts. This will allow us to deep our knowledge on the evolution of the luminous and the dark matter distributions obtaining decisive evidences about the actual DM scenario. 
 \end{enumerate}

Finally, it goes without saying that a large flux of observations will come from measurements from radio telescopes as ALMA and SKA and from optical (near infrared/visible light) telescopes as WFIRST and ELT.

\vspace{6pt}
\authorcontributions{The authors contributed equally to this work. All authors have read and agreed to the published version of the manuscript.}

\funding{This research received no external funding. }

\acknowledgments{\textls[-15]{We thank Fabrizio Nesti, Nicola Turini and Andrea Lapi for very useful discussions.}}

\conflictsofinterest{The authors declare no conflict of interest.}

\end{paracol}
 
\begin{adjustwidth}{+0.1cm}{0cm}
\printendnotes
\end{adjustwidth}

\reftitle{References}

\end{document}